\documentclass[reprint,superscriptaddress,
aip,rsi,
amsmath,amssymb]{revtex4-1}
\usepackage{graphicx}
\usepackage{dcolumn}
\usepackage{bm}
\usepackage{color}
\usepackage{graphicx}
\usepackage{hyperref}
\usepackage{amsmath}
\usepackage{titlesec} 
\titlespacing\section{0pt}{10pt}{4pt}
\titlespacing\subsection{0pt}{10pt}{2pt}

\newcommand{\thetaF}{\theta_{_{\text{F}}}}

\begin{document}
\title{Continuous-Wave Cavity Ring-Down Polarimetry}
\author{Jim C. Visschers}
\affiliation{Institut f\"ur Physik, Johannes Gutenberg Universit\"at-Mainz, 55128 Mainz, Germany}
\author{Oleg Tretiak}
\affiliation{Institut f\"ur Physik, Johannes Gutenberg Universit\"at-Mainz, 55128 Mainz, Germany}
\author{Dmitry Budker}
\affiliation{Institut f\"ur Physik, Johannes Gutenberg Universit\"at-Mainz, 55128 Mainz, Germany}
\affiliation{Helmholtz-Institut Mainz, Mainz 55128, Germany}
\affiliation{Department of Physics, University of California, Berkeley, California 94720-300, USA}
\author{Lykourgos Bougas}\email{lybougas@uni-mainz.de}
\affiliation{Institut f\"ur Physik, Johannes Gutenberg Universit\"at-Mainz, 55128 Mainz, Germany}
\date{\today}

\begin{abstract}
We present a new cavity-based polarimetric scheme for highly sensitive and time-resolved measurements of birefringence and dichroism, linear and circular, that employs rapidly-pulsed single-frequency CW laser sources and extends current cavity-based spectropolarimetric techniques. We demonstrate how the use of a CW laser source allows for gains in spectral resolution, signal intensity and data acquisition rate compared to traditional pulsed-based cavity ring-down polarimetry (CRDP). We discuss a particular CW-CRDP modality that is different from intensity-based cavity-enhanced polarimetric schemes as it relies on the determination of the polarization-rotation frequency during a ring-down event generated by large intracavity polarization anisotropies. We present the principles of CW-CRDP and validate the applicability of this technique for measurement of the non-resonant Faraday effect in solid SiO$_2$ and CeF$_3$ and gaseous butane. We give a general analysis of the fundamental sensitivity limits for CRDP techniques and show how the presented frequency-based methodology alleviates the requirement for high finesse cavities to achieve high polarimetric sensitivities, and, thus, allows for the extension of cavity-based polarimetric schemes into different spectral regimes but most importantly renders the CW-CRDP methodology particularly suitable for robust portable polarimetric instrumentations.
\end{abstract}

\maketitle
\tableofcontents

\section{Introduction}
Polarimetry is one of the oldest research tools available to characterize the optical properties of a substance, and has been essential in scientific developments, such as, for identifying a molecule's structure\,\cite{Eliel2008,KELLY2005}, in tests of fundamental symmetries of nature\,\cite{FORTSON1984}, in optical magnetometry\,\cite{budkerBook}. Polarimetry is also used in industrial applications where it is routinely applied in quality and process control in the pharmaceutical, chemical, and agricultural industries\,\cite{Busch2006}. Despite their extensive use, the sensitivity limits of commercially available optical spectro-polarimeters are at the $\sim10\,\mu$rad levels, which constrain the applicability of polarimetry in a wide range of important research and industrial applications, e.g., in chemical analysis and drug design and development where the relatively poor polarimetric sensitivities result in poor analyte concentration detection limits at the micromolar levels\,\cite{KELLY2005,ARGENTINE2007}, or in trace-gas detection and analysis.\\ 
\indent Different techniques have been developed to push optical polarimetry to its limits that, in principle, are fundamentally constrained by the photon shot-noise of the light source used for the measurements, which for a $\sim1$\,mW of visible radiation are at the $\sim\!10^{-8}$\,rad/$\sqrt{\rm{Hz}}$ levels\,\cite{Budker2008}. However, there exist different sources of noise (e.g. light-intensity noise and detector electronic noise) that are usually much larger than the photon shot noise and limit polarimetric sensitivities to the $10^{-7}-10^{-6}$\,rad/$\sqrt{\rm{Hz}}$ levels. Modulation techniques are typically employed to approach shot-noise limited polarimetric sensitivities, but even then these limits preclude broad application of polarimetry, especially in trace-gas detection and analysis. The most straightforward solution towards increased spectropolarimetric sensitivities is through the enhancement of the interaction path-length of the probing radiation in the substance under investigation. This can be achieved with the use of optical mirrors to create either multipass cells or optical cavities, where in both cases spectropolarimetric signals are enhanced by the average number of passes through the medium. Multipass cells\,\cite{Li2011,Tuzson2013}, such as White cells\,\cite{White1942} or Herriott cells\,\cite{Herriott1964}, are technically easy to construct and implement, and path-length enhancements as large as $\times$500 have been demonstrated\,\cite{Das2011,Krzempek2013}. However, multipass-based techniques are still limited by laser intensity fluctuations and the probing radiation travels along a different path for each pass, and, thus, relatively large substance volumes are typically required, making, for example, multipass cells hard to implement for measurements of liquid samples. Importantly, multipass techniques are suitable for absorption measurements and, in principle, can not be employed for the measurement of natural optical activity. On the contrary, using stable optical cavities one can achieve path-length enhancements of up to $10^5$ using state-of-the-art high quality mirrors, with effective path-lengths of up to several hundred km (compare this to the 10\,cm interaction path-length of a single-pass commercial polarimeter), enabling record sensitivities for measurements of absorption and birefringence. Moreover, optical cavities can be easily made compact and allow for light-medium interactions in small volumes.\\
\indent There exist several cavity-based polarimetric designs for the measurement of linear and circular, birefringence and dichroism. Although there are fundamental symmetry considerations for which type of cavity one employs for the measurement of reciprocal or non-reciprocal birefringent effects\,\cite{barron_2004,Mueller2000,bougas2012}, in general, stable optical cavities consist of two to four optical mirrors, and all designs can be realized using (a) continuous-wave (CW) laser light or (b) short laser pulses. In the case of using CW laser sources, state-of-the-art polarimetric sensitivities of $10^{-13}$\,rad$\sqrt{\rm{Hz}}$ have been demonstrated\,\cite{Durand2010}. There exist several approaches for performing CW cavity-based polarimetric measurements, including modulation-based ones (e.g. Ref.\,\cite{Gianella2017}), or techniques developed in the field of frequency metrology\,\cite{Hall2000,Gilles2010}. However, these CW-laser-based cavity-enhanced polarimetric techniques require complicated and extremely stable optomechanical setups together with state-of-the-art electronics to achieve this kind of sensitivity levels, precluding their translation into portable instrumentation and, therefore, for applicability in field studies. \\
\indent In the case of using short laser pulses, cavity-based polarimetric techniques build upon the inherent sensitivity of cavity ring-down spectroscopy (CRDS)\,\cite{Wheeler1998,Berden2009}, via the insertion of polarization-selective components into the preparation, interaction, and analysis stage of a conventional CRDS apparatus. Particularly, in CRDS a laser pulse is stored in a stable optical cavity containing a sample, and the pulse decay, characterized by the ``ring-down" time $\tau$, is monitored with the aid of a detector that measures the intensity of light transmitted through the mirrors. Sensitive detection of this decay time, which depends on the intracavity absorption losses, allows for highly sensitive absorption measurements that are inherently insensitive to the intensity noise originating from the light source. For the case of pulsed polarimetry we can distinguish up to date two distinct measurement approaches. The first is based on monitoring polarization-dependent changes in $\tau$ as the frequency of the laser source is scanned across a transition of the system under investigation that demonstrates dichroism (linear or circular). The system's dichroism (and its associated birefringence) will transfer a fraction of the intra-cavity optical power from one of its eigenpolarization states to its orthogonal one, which thus becomes observable through the losses measured by the ring-down decays on two orthogonal polarization-sensitive detection channels. Such measurements are effectively intensity-based and can be categorized under the general term of polarization-dependent cavity ring-down spectroscopy (PD-CRDS), a terminology introduced in the first demonstration by Engeln and co-workers\,\cite{Engeln1997,Engeln1998}. PD-CRDS schemes have achieved near-shot-noise limited sensitivities at $\!<\!10^{-9}$\,rad$\sqrt{\rm{Hz}}$ levels, and have been employed for measurements of linear and non-reciprocal circular birefringent effects, such as the measurement of residual or stress-induced linear birefringence of supermirrors\,\cite{YongLee2000,Huang2008,Dupre2015,Fleisher2016}, and the measurement of resonant Faraday optical rotation (otherwise known as FR-CRDS) as employed for the sensitive and selective detection of paramagnetic gaseous molecules and radicals (e.g. O$_2$\,\cite{Berden1998,Hayden2018,Patrick2019,Gianella2019}, HO$_2$\,\cite{Gianella2018,Gianella2019}). However, PD-CRDS is specifically suitable for resonant effects and requires ultrahigh finesse optical cavities (as the sensitivity in the decay time is translated directly into polarimetric sensitivity), which effectively increases the cost of the apparatus and can limit its operational lifetime (degradation of the high quality optical mirrors results in reduced sensitivities), precluding this approach from being the most suitable one for portable instrumentations for field studies. \\
\indent The second approach, alternative to PD-CRDS, relies on the use of an intracavity ``bias" polarization anisotropy that introduces a large background polarization rotation and results in a rapid oscillating signal superimposed upon the ring-down signal as detected in a polarization-sensitive analysis stage. This polarization bias rotation offers two critical advantages. The first one is to provide an easily measured polarization beat frequency that is altered with the addition of an incremental anisotropy (i.e. an effect under investigation), transforming the polarimetric measurement into a frequency-based one. For this reason, we categorize these techniques under the general term of cavity ring-down polarimetry (CRDP). The second advantage is that the large bias polarization rotation can suppress intracavity anisotropies of opposite symmetry that might otherwise affect the desired measurement. This advantage becomes apparent in the case of chiral sensing, where CRDP approaches have been implemented for highly sensitive chiral polarimetry\,\cite{Mueller2000,Muller2002,Sofikitis2014,Bougas2015}. Particularly, in the case of chiral sensing, any intracavity linear birefringence effect would otherwise inhibit the sensitive measurement of the expected weak chiroptical signals. The solution is to use a large intracavity circular birefringence, much larger than any residual linear birefringence present within the cavity, which introduces a large background polarization rotation and protects the measurement of the weak chiroptical signals. Despite the impressive successes of (chiral-sensitive) CRDP\,\cite{Sofikitis2014}, the demonstrated sensitivities are still several orders of magnitude worse than their expected shot-noise limits.\\
\indent Most of the PD-CRDS demonstrations have employed CW laser sources, while all CRDP demonstrations have employed thus far only pulsed laser sources. The extension in using CW laser sources in CRDS was first shown by Romanini and co-workers, who demonstrated how CW-based CRDS allows for gains in spectral resolution, signal intensity, and data acquisition rates\,\cite{ROMANINI1997}. In CW-CRDS, the frequency of the CW laser is either locked into resonance with the CRD cavity or brought into resonance slowly (at rates sufficient slow to allow for maximum input coupling, constrained by the cavity lifetime), and a ring-down signal is observed after the laser beam is quickly interrupted (typically much faster than the cavity lifetime, i.e. $\tau$). This procedure guarantees a substantial buildup of the intracavity field leading to strong ring-down signals, as the signal intensity of the transmitted beam can reach close to the input intensity, enabling photon-shot-noise limited CRDS measurements. Several commercially available portable systems based on diode-laser CW-CRDS already exist\,\footnote{Los Gatos Research, http://www.lgrinc.com; Picarro, https://www.picarro.com}, however none yet for highly sensitive polarimetry.\\
\indent In this article, we present the technique of CW-CRDP, a methodology that combines the strengths of all the aforementioned techniques and alleviates their weaknesses, and which we consider to be the ideal modality for highly sensitive, cost-effective, polarimetry, particularly suitable for portable and robust spectro-polarimetric instrumentation. CW-CRDP extends the operational principles of pulsed-based CRDP by employing CW-laser sources, enabling time-resolved shot-noise limited spectropolarimetric sensitivities.\\
\indent We start by presenting the general principles of operation of CW-CRDP and discuss how these can be generally adapted for measurements of birefringence and dichroism, both circular and linear. Our proposed methodology relies on the practice of using a bias intracavity polarization anisotropy as a means of protection of weak polarimetric signals under investigation, and while this approach has already been applied several times in cavity-based polarimetry, particularly in the case of studying chirality\,\cite{Mueller2000,Muller2002,Sofikitis2014,Bougas2015}, the fundamental measurement sensitivity benefits behind this methodology have not been explored in detail before, and we demonstrate here how CW-CRDP can reveal these. \\
\indent In particular, in Sec.\,II we present the theory behind CW-CRD polarimetry, and its operational principles. In Sec.\,III, we present the details of a prototype setup that we employ for measurements of non-resonant Faraday optical rotation from solid and gaseous systems using CW-CRDP. In Sec.\,IV we present results for the Faraday effect in solid SiO$_2$ and CeF$_3$, and in gaseous butane, where we measure their Verdet constants to quantitatively validate our method. Finally, in Sec.\,V we discuss the fundamental sensitivity limits for any CW-CRD-based polarimetric scheme, which we employ to analyze the stability of our experimental results. Most importantly, we demonstrate how the introduction of an intracavity bias anisotropy, which enables frequency-based polarimetric measurements and is adaptable in all CRD-based polarimetric schemes, can effectively alleviate the necessity for high finesse optical cavities and, therefore, the need of ultrahigh-quality optics, which can be costly and not available at all optical spectral regions, a crucially important aspect for robust and versatile portable spectropolarimetric instrumentations.

\section{Theory}
We start by discussing the theory of CW-CRDP. Although several extended analyses of the theory behind CW-based cavity-enhanced polarimetric schemes exist in the literature\,\cite{Nilsson1989,VALLET1999,Bougas2014,Gianella2019}, we repeat here key theoretical aspects to clarify the necessary terminology and concepts required to understand the experimental principles of the CW-CRDP technique.\\
\indent To examine the principles of operation of a CW-CRD polarimetric protocol we focus, for simplicity and without loss of generality, on linear optical cavities, consisting of two mirrors and, as a bias intracavity polarization anisotropy, a non-reciprocal circular birefringent effect (e.g. Faraday effect). However, we emphasize that the general theoretical approach we present here can be similarly applied for the case of reciprocal circular birefringence (e.g. for a four-mirror cavity for the study of chirality\,\cite{Bougas2014}) and linear birefringence (e.g. linear cavity for the study of Voigt effect, mirror-related birefringence)\,\cite{Engeln1997,Dupre2015}, and, when required, we discuss extensions to different designs. Furthermore, we assume that the laser beam is mode-matched into the TEM$_{00}$ mode of the optical cavity and, therefore, we focus our analysis on the polarization properties of the cavity's longitudinal modes. We also neglect any changes in the spatial profile of the laser beam, possibly introduced by the intracavity element(s). \\
\indent We present an eigenpolarization theory for the cavity-based polarimeter based on the Jones matrix calculus that allows us to describe the full optical system and to incorporate a CW laser source and the ability to switch it on and off rapidly. Here, the full optical system includes the polarization-control optics before the cavity, the optical cavity including the intracavity anisotropy, and the polarization-analysis (detection) stage. In the Jones matrix formalism, the effect of any optical element on the polarization state vector of the laser light is described as a linear operator expressed by a $2\times2$ matrix whose elements are in general complex. We denote each of these matrices by boldface letters $\mathbf{J}$. Most importantly, the direct incorporation of amplitude and phase information in the Jones matrices allows for the investigation of coherent phenomena.

\subsection{Jones matrices for polarization optics}
The Jones matrix representation of a mirror with a finite reflectivity $R$ ($0\leqslant R<1$) is given as:
\begin{equation}
\mathbf{J}_{{\rm M}_i}(R_i,\delta_i)=\sqrt{R_i} \left(\begin{array}{cc}-e^{i\delta_i/2} & 0 \\0 & e^{-i\delta_i/2} \end{array}\right),
\end{equation}
where the index $i$ denotes each cavity mirror (in our case $i$=$\{1,2\}$). Since we focus on two-mirror cavities, with normal angle-of-incidence reflections, we set the Fresnel amplitude reflection coefficients for the \textit{s} and \textit{p} polarizations to be equal in magnitude (an assumption expressed by the common factor $\sqrt{R_i}$). The differential $s$-$p$ phase shift $\delta_{i}\equiv\delta^i_p-\delta^i_s$, represents the linear birefringence obtained upon mirror reflection. For normal incidence these $s$-$p$ phase shifts can be of the order of $10^{-7}-10^{-5}$\,rad at a specific design wavelength\,\cite{Huang2008,Hall2000}.\\
\indent In the presence of a longitudinal magnetic field a medium becomes circular birefringent, an effect otherwise know as the Faraday effect\,\cite{BudkerRMP}. The Faraday optical rotation is expressed as: $\thetaF=V\!\cdot\!\text{B}\!\cdot\! l$, where B is the magnetic field strength along the direction of light propagation, $l$ is the path-length in the medium, and $V$ is the Verdet constant of the medium. The Jones matrix representation for the Faraday effect is an SU(2) rotation matrix with argument $\theta_{_{\text{F}}}$:
\begin{equation}
\mathbf{J}_{_{\text{F}}}(\theta_{_{\text{F}}})=\left(\begin{array}{cc} \cos\theta_{_{\text{F}}}& -\sin\theta_{_{\text{F}}} \\ \sin\theta_{_{\text{F}}} & \cos\theta_{_{\text{F}}} \end{array}\right).
\end{equation}
The physical direction of the polarization rotation is defined by the magnetic field orientation. Due to the non-reciprocal nature of the Faraday effect, when either the magnetic field or the direction of propagation of the light reverses, the sign of rotation reverses. This directional symmetry breaking, induced by the Faraday effect, has been essential for the implementation of crucial signal reversals in chiral cavity-based polarimetry\,\cite{Sofikitis2014,Bougas2015}.

\subsection{Frequencies and polarizations of the cavity spectrum}
\begin{figure}[t!]
    \includegraphics[width =0.8\linewidth]{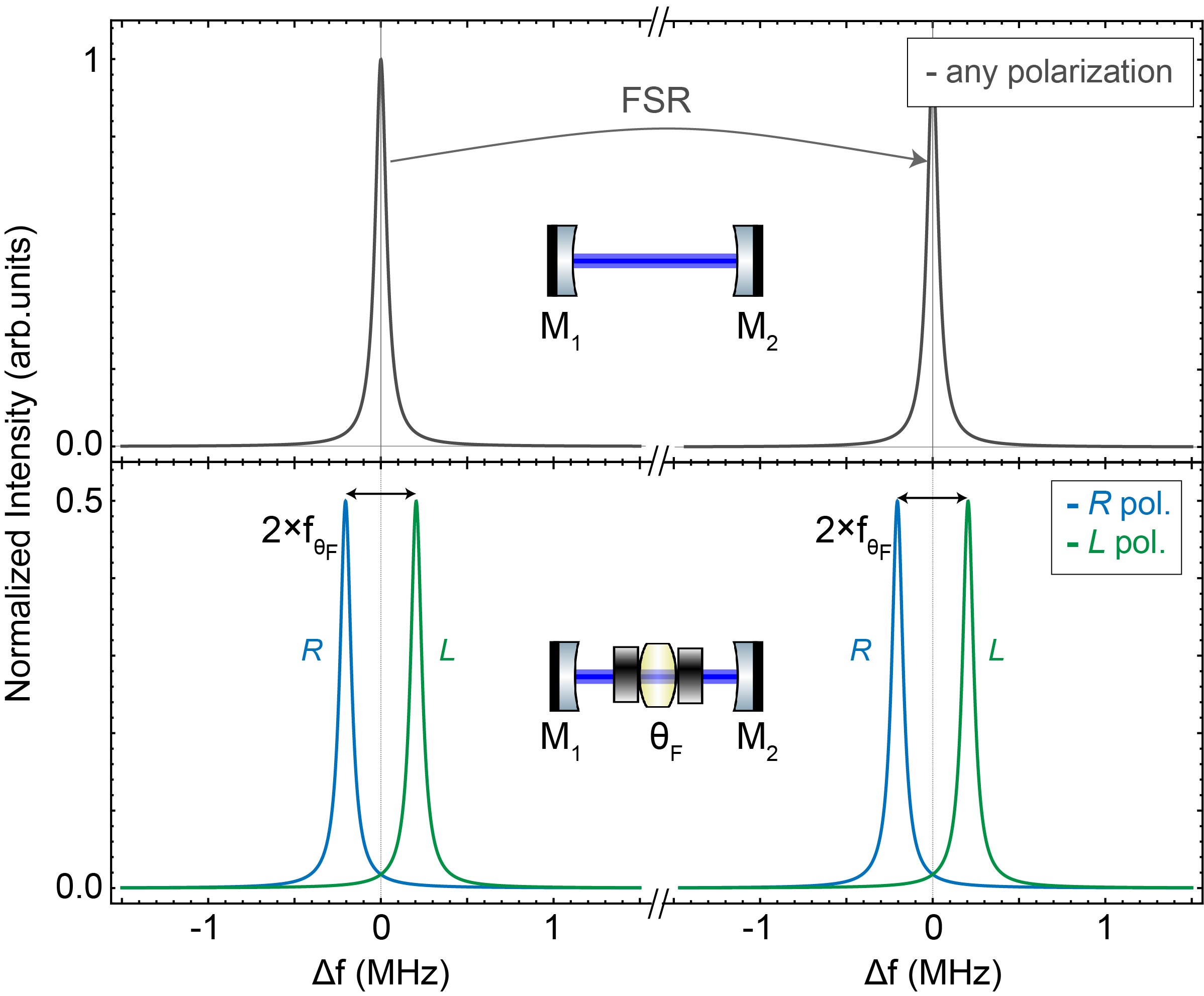}
    \caption{\small{Cavity frequency polarization spectrum of a linear (two-mirror) optical cavity in the (a) absence and (b) in the presence of an intracavity polarization anisotropy. We choose here as an intracavity anisotropy a non-reciprocal Faraday effect, which splits the cavity spectrum into two (orthogonal) circularly polarized modes, $R$ and $L$, by 2\,f$_{\thetaF} = 2\,\thetaF\times(\pi/{\rm{FSR}})$, where FSR is the cavity's free spectral range (for clarity we assume here a $\thetaF$ value much larger than the cavity linewidth).}}
    \label{fig:CavModes}
\end{figure}

The Jones matrix representation of a round-trip in the optical cavity is obtained by the ordered multiplication of Jones matrices representing the independent optical elements (see Fig.\,\ref{fig:CavModes}),
\begin{equation}
\mathbf{J}^{_{\text{cav}}} =  \mathbf{J}_{{\rm M}_1}(R,\delta) \!\cdot\!  \mathbf{J}_{_{\text{F}}}(-\theta_{_{\text{F}}}) \!\cdot\!  \mathbf{J}_{{\rm M}_2}(R,\delta) \!\cdot\!  \mathbf{J}_{_{\text{F}}}(\theta_{_{\text{F}}}) ,
\label{eq:Jcavity}
\end{equation}
where, again for simplicity, we assume that mirrors M$_1$ and M$_2$ have the same characteristics, i.e. $R_1=R_2\equiv R$ and $\delta_1=\delta_2\equiv\delta$.\\
\indent Using Eq.\,\ref{eq:Jcavity} we can determine the allowed polarizations of the cavity modes (eigenpolarizations) along with their respective frequencies. Considering that the Jones matrices are unitary matrices of rank two, each matrix has two eigenvalues and two eigenvectors; the eigenvectors $\nu_{\pm}$ are orthogonal, in general complex, vectors and represent the eigenpolarizations of each cavity mode. The eigenvalues can be written in the general form $\lambda_{\pm}=e^{\pm i\alpha}$, and the phase of each eigenvalue, $\alpha$, is the round-trip optical phase shift obtained during light propagation, which, thus, yields the frequency splittings of the eigenmodes. \\
\indent In the simple case of an isotropic cavity ($\theta_{_{\text{F}}}\!=\!0$ \& $\delta\!=\!0$), the two eigenmodes are degenerate and any polarization state can couple into the cavity (i.e. $\mathbf{J}^{_{\text{cav}}}$ becomes proportional to the identity matrix; Fig.\,\ref{fig:CavModes}, upper panel). The introduction of any polarization anisotropy lifts this degeneracy. In the most general case, the spectrum of the cavity is represented by two non-degenerate modes of elliptical polarization, whose frequencies lie above and below the degenerate frequency of the isotropic case. However, when the intracavity anisotropy is a circular birefringence ($\theta_{_{\text{F}}}\!\neq\!0$ \& $\delta\!=\!0$), the two eigenpolarization modes are circular-polarization states, denoted hereafter as $R$ and $L$ modes, and, thus, the cavity spectrum is now represented by two modes split in frequency by 2\,f$_{\thetaF}=2\theta_{\rm{F}}\cdot\rm{FSR}/\pi$, where FSR$=(c/L_{rt})$ is the cavity's free spectral range, with $c$ the speed of light and $L_{rt}$ the round-trip cavity length (Fig.\,\ref{fig:CavModes}, lower panel). We note that this mode structure can be resolved in the case of a frequency splitting much larger than the cavity linewidth, which is the case when the polarization anisotropy (here a Faraday effect), obeys the relationship $\thetaF\gg\pi/\mathcal{F}$, where $\mathcal{F}$ is the finesse of the optical cavity. \\
\begin{figure*}[t!]
    \includegraphics[width=0.95\linewidth]{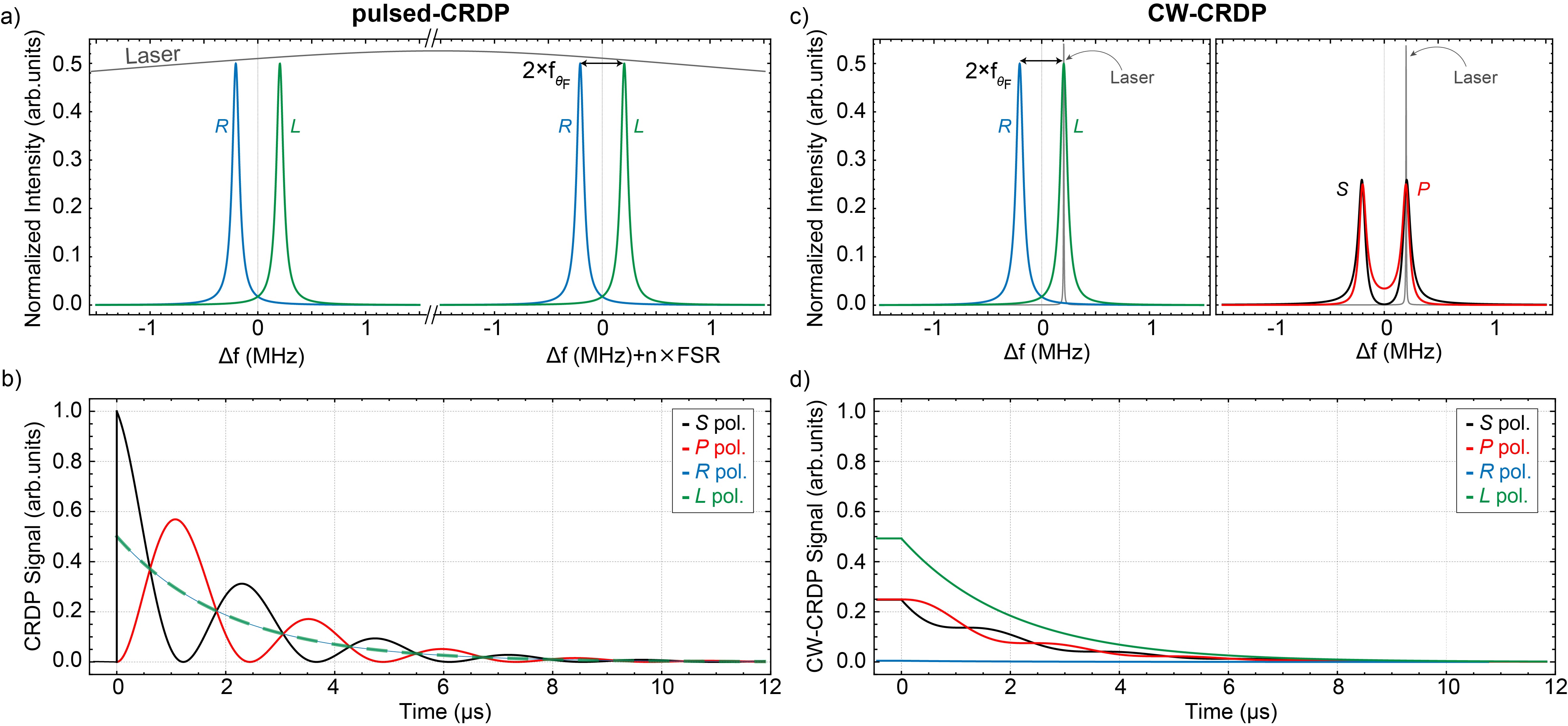}
    \caption{\small{\emph{Principles of CRDP measurements -} (a)\,\&\,(b) pulsed-CRDP, (c)\,\&\,(d) CW-CRDP. For the simulations we assume a two-mirror cavity with an intracavity circular birefringence, i.e. Faraday effect (we use cavity length of 0.6\,m, cavity finesse of $\mathcal{F}\approx3140$, and $\thetaF=2.6$\,mrad). Typically, the frequency linewidth of a pulsed laser is much larger than several ($n$) cavity FSRs, which allows for the direct coupling of an incident linearly polarized beam into the cavity [(a)]. The transmitted radiation is analyzed with a linear balanced polarimeter yielding a ring-down signal superimposed with a polarization frequency beating that has a modulation depth approaching unity, i.e. the CRDP trace (when the circular components are analyzed, pure exponentially decaying signals are observed demonstrating the equal coupling to both $R$\,\&\,$L$ modes) [(b)]. In contrast, in CW-CRDP the linewidth of the laser is typically much smaller than a cavity linewidth and the incident radiation couples into a single cavity mode [for this reason, in (c), we show only one pair of $R$\,\&\,$L$ modes, instead of multiple mode pairs separated by several FSRs as in (a)]. Depending on the detuning with respect to the $R$-$L$ cavity modes, different polarization states are allowed to couple into the cavity [(c)]. Once the incident (and frequency locked) radiation is switched off, at time $t\!=\!0$, the excited polarization mode evolves resulting in pure exponentially decaying signals when analyzed with a circularly sensitive balanced polarimeter. However, when analyzed with a linearly sensitive balanced polarimeter, which appropriately mixes the circular polarization components of the transmitted radiation to produce two orthogonal, linearly polarized ($S$ and $P$) waves, one observes ring-down signals superimposed with a polarization beat frequency that now has a modulation depth and amplitude defined by the initially excited polarization state [(d)]. In (a)\,\&(c) the gray dotted lines at zero detuning with respect to each $R$-$L$ mode pair (i.e. $\Delta{\rm{f}}=0$) correspond to the two-fold degenerate axial mode of an isotropic cavity. }}
    \label{fig:Principles}
\end{figure*}
\indent In the case of intracavity linear birefringence ($\theta_{_{\text{F}}}\!\neq\!0$ \& $\delta\!\neq\!0$), which can also originate from thermal or mechanical stress in all intracavity optics apart from mirror-reflection-related phase shifts, the path-length-related enhancement of circular birefringence is inhibited through the transformation of the circularly polarized eigenmodes into linear ones. The equivalent time-domain explanation is that an incident linearly polarized light beam will oscillate between linear and circular polarization states, reducing the effective path-length enhancement and the sensitivity of the measurement. However, as long as the bias intracavity circular birefringence is much larger than linear birefringence, i.e. here $\thetaF\gg\delta$, then the effects of linear birefringence will be averaged out and the cavity modes will maintain the circular polarization character. The physics behind this process has been extensively discussed within the context of chiral-sensitive CRDP\,\cite{Muller2002,Bougas2014,SOFIKITIS2018}.\\
\indent Finally, in the case of circular dichroism, the linewidths of the two cavity eigenpolarization modes become different, since the cavity finesse depends on the intracavity losses, which are in this case different for the two circular polarization states.

\subsection{Principles of a CRDP measurement: pulsed vs. CW} 
In traditional CRDP, laser pulses with linewidths much larger than a cavity FSR are used [Fig.\,\ref{fig:Principles}\,(a)]. For example, a gaussian-shaped Fourier-limited laser pulse of $\sim\!1$\,ps in duration has a spectral bandwidth of $\sim\!0.4$\,THz, which is much larger than the FSR of meter-long cavities. Typically, such a linearly polarized laser pulse is directly coupled into the cavity by coherently exciting multiple cavity (eigenpolarization) modes. The transmitted beam is also linearly polarized, and when analyzed using a linear-polarization detector, i.e. a balanced polarimeter that analyzes linear polarization components such as a Wollaston prism, the resulting ring-down signal is superimposed with the differential polarization beat frequency equal to the $R$-$L$ frequency mode-splitting [CRDP trace: Fig.\,\ref{fig:Principles}\,(b); note that if we analyze the circular components we see pure exponentially decaying ring-down signals]. If there are no depolarization mechanisms, such as dichroic absorption losses and/or any residual intracavity linear birefringence, the modulation depth of this polarization beat frequency approaches unity.\\
\begin{figure}[t!]
    \includegraphics[width=\linewidth]{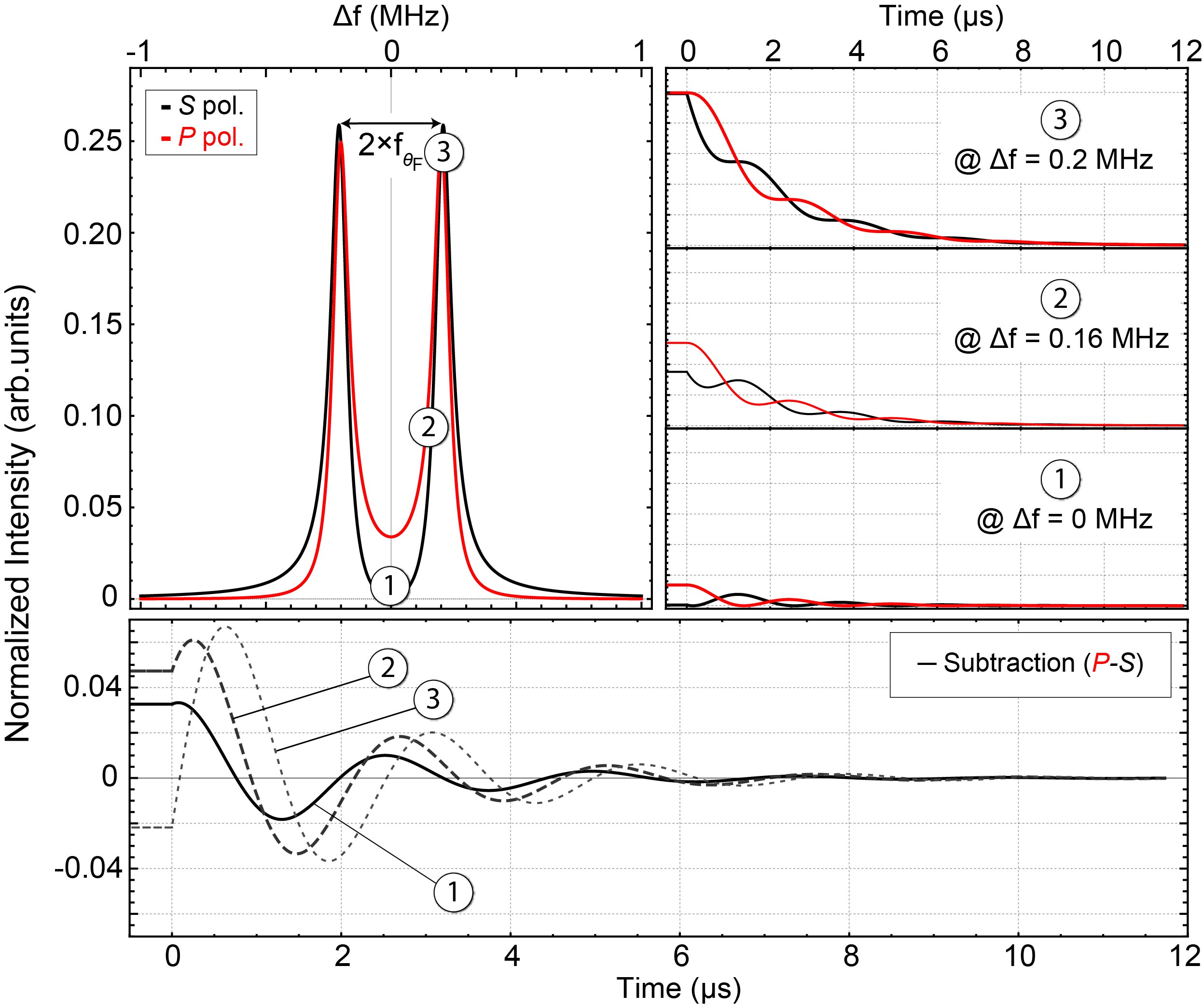}
    \caption{\small{Depending on the detuning of the frequency of the incident radiation with respect to the $R$ and $L$ cavity eingepolarization modes, a different polarization state is coupled into the cavity, and once the incident (and frequency locked) radiation is switched off at time $t\!=\!0$, the excited polarization mode evolves resulting in a ring-down signal superimposed with a polarization beat frequency, as recorded by the $S$ and $P$ polarization channels of a linear polarimeter. The amplitude, modulation depth, and phase of the resulting CW-CRDP traces depend on the frequency detuning (upper right panels). Obtaining their difference via a balanced detection scheme, yields an exponentially decaying sinusoidal signal where one can clearly observe the amplitude and phase differences between the resulting CW-CRDP traces (lower panel). For the simulations we use the same parameters as for Fig.\,\ref{fig:CavModes} and Fig.\,\ref{fig:Principles}, and the normalized signals are with respect to the maximum intensity one could achieve in an empty cavity (Fig.\,\ref{fig:CavModes}). }}
    \label{fig:Split}
\end{figure}
\indent In the case of CW-CRDP, the measurement principle is different. Incident radiation from a CW laser source with a linewidth that is significantly smaller than that of a cavity, can couple into and build-up within that cavity by exciting only one particular cavity mode at a time [Fig.\,\ref{fig:Principles}\,(c); note that here we show only one pair of $R$\,\&\,$L$ modes, instead of multiple mode pairs separated by several FSRs as in Fig.\,\ref{fig:Principles}\,(a)]. As shown in the previous section, for a linear cavity with a large intracavity circular birefringent anisotropy, the cavity eigenpolarization modes are frequency-separated circularly polarized modes ($R$ and $L$ modes), and the polarization state of the incident radiation will be projected onto these. Therefore, depending on the frequency detuning of the incident radiation with respect to the $R$-$L$ cavity eingepolarization modes, a different polarization state will be coupled into the cavity. For example, the frequency at which the $R$ and $L$ modes are equally excited [$\Delta {\rm{f}}\!=\!0$\,MHz; Fig.\,\ref{fig:Principles}\,(c)] corresponds to a mode of linear polarization, as this is an equal coherent superposition of right and left circularly polarized states. Similarly, the central frequencies of the individual $R$ and $L$ cavity modes correspond to elliptical polarization states (since these are an unequal superposition of the $R$ and $L$ modes), and the degree of ellipticity of these states depends on the detuning but also on the $R$-$L$ mode-splitting, i.e. on the value of the bias anisotropy (e.g. $\thetaF$), as the larger (smaller) the splitting, the larger the circular (linear) character of the coupled, and consequently of the transmitted, polarization state. \\
\indent As in pulsed-CRDP, in CW-CRDP one analyzes the polarization state of the transmitted radiation using a linearly-sensitive balanced polarimeter, which appropriately mixes the circular polarization components of the transmitted radiation to produce two orthogonal, linearly polarized ($S$ and $P$) waves incident upon separate photodetectors [Fig.\,\ref{fig:Principles}\,(c)]. However, in CW-CRDP, to initiate a ring-down event, the incident radiation which is coupled into and built-up within the cavity, is abruptly switched off (at time-scales much faster than the cavity lifetime). Since the frequency detuning of the incident radiation defines the coupled polarization state (which can vary continuously from linear to circular), once the incident radiation is switched off, this coupled (intracavity) polarization state freely evolves, resulting in a transmitted ring-down signal superimposed with a polarization beat frequency [CW-CRDP trace: Fig.\,\ref{fig:Principles}\,(d)]. For the case of incident radiation tuned to the central frequency point ($\Delta {\rm{f}}\!=\!0$\,MHz), linearly polarized light couples to and is transmitted by the cavity, which results in a modulation depth that reaches unity, similarly to the case of pulsed-CRDP. However, in the CW case, the signal amplitude is reduced, and this reduction follows the value of the intracavity polarization anisotropy (i.e. the $R$-$L$ mode-splitting). Similarly, the circularity of the coupled light increases as the incident radiation is detuned from this central frequency point, leading to a decrease in the modulation depth of the polarization beat frequency,  though accompanied by an increase in amplitude. We demonstrate these cases in Fig.\,\ref{fig:Split} where we present simulations of CW-CRDP traces for different possible frequency detunings of the incident radiation with respect to the $R$-$L$ modes (Fig.\,\ref{fig:Principles}). Overall, to perform CW-CRDP measurements using narrow-linewidth laser sources, an optimum frequency detuning of the incident radiation with respect to the cavity's eigenpolarization modes exists that depends on the cavity linewidth (i.e. the cavity finesse) and the strength of the intracavity anisotropy (e.g. $\thetaF$). For example, if the strength of the intracavity anisotropy is large enough to result in a mode splitting much larger than a few cavity linewidths, the two modes are practically decoupled and it is not be possible to perform measurements using a narrow-linewidth laser (or, equivalently, if the finesse of the cavity is high enough to effectively lead to a substantially reduced mode overlap). On the other hand, if the mode splitting is much smaller than a cavity linewidth, then substantial overlap between the $R$-$L$ modes is attained and one achieves optimum overlap for a large range of frequency detunings. However, is such case, the polarization beat frequency signal becomes small, which might lead to reduced polarimetric sensitivities (e.g. because the intracavity circular birefringence is comparable in strength to any residual intracavity linear birefringence). In Sec.\,V we discuss how one should appropriately select the experimental parameters of a CW-CRDP technique when we analyze the fundamental sensitivity limits of CRDP methods, and alternative experimental modalities that resolve issues related to the mode coupling of the incident radiation. \\
\indent In Fig.\,\ref{fig:Split} we also show that subtracting the $P$ and $S$ signals generates a pure damped sinusoidal signal which clearly exhibits the change in signal amplitude as a function of the frequency detuning of the incident radiation with respect to the cavity's eigenpolarization mode-splitting. In a CW-CRDP protocol, a balanced detection scheme is advantageous as the signal subtraction suppresses common mode laser intensity noise measured by the two $S$- and $P$-wave photodetectors, resulting in improved signal-to-noise ratios. Furthermore, in addition to the amplitude change as a function of detuning, our simulations show an associated phase shift in the recorded beat frequency. This phase shift dependence on the frequency detuning is critical for a CW-CRDP protocol because it indicates that the performance of a frequency-stabilization scheme, which controls the frequency of the incident radiation with respect to the cavity-mode structure, directly impacts the measurement sensitivity (in pulsed-CRDP this is not an issue as the incident radiation couples fully with multiple modes). We discuss this in more detail when we analyze the fundamental sensitivity limits of the CW-CRDP method.\\
\indent Here, we wish to emphasize a critical difference between pulsed-based and CW-based CRDP. In the case of CW-CRDP the initial signal amplitude is defined by the cavity transmission on resonance, which in theory can reach unity, so detected optical powers of 1-100\,mW can be achieved, whereas for pulsed-CRDP typical detected optical powers are at the $\mu$W levels at best\,\cite{ROMANINI1997,Sofikitis2014}. Therefore, CW-CRDP can result in signals with significantly higher signal-to-noise ratio than pulsed techniques, while attaining much higher repetition rates through the use of fast optical switches, leading to improved (ideally shot-noise limited) statistics for a given integration time. \\
\indent Finally, although this discussion focuses on measurements of circular birefringence, the same measurement principles are applicable in studies of linear birefringence. Moreover, the case of linear (circular) dichroism is studied by measuring the ring-down times as recorded by a linear (circular) balanced polarimeter.

\section{Experiment}
\subsection{CW-CRDP apparatus}
\begin{figure}[t!]
    \includegraphics[width=\linewidth]{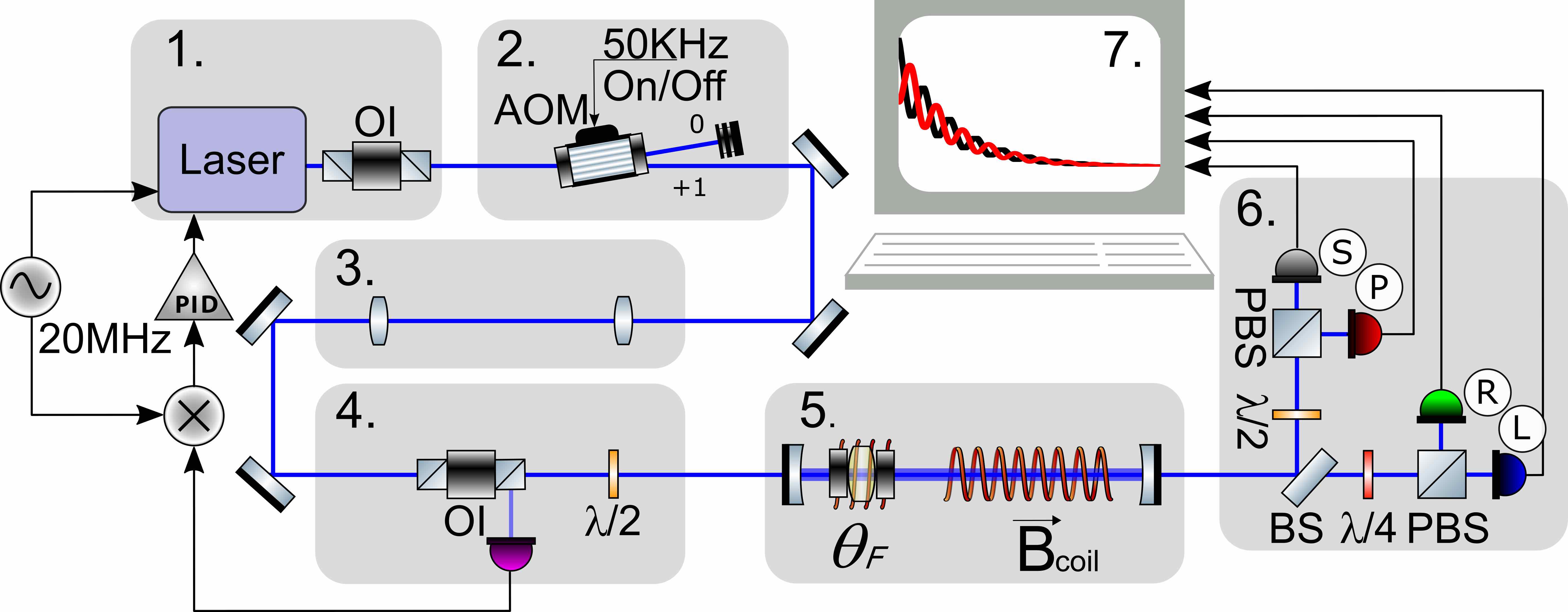}
    \caption{\small{Schematic diagram of the CW-CRDP optical setup: 1) Laser operating in continuous-wave mode followed by an optical isolator (OI) that blocks unwanted back-reflections to the laser source; 2) a fast optical switch (acousto-optic modulator; AOM) is used to rapidly interrupt the laser beam and initiate a ring-down event; 3) beam-shaping optics to enable optimal coupling of the laser beam into the optical cavity; 4) polarization preparation stage and optoelectronics required to generate an error signal; 5) two-mirror optical cavity with an intracavity Faraday anisotropy, characterized by its rotational strength, $\thetaF$, and permanent magnets and coils for the generation of axial magnetic fields; 6) light transmitted through the cavity is split using a beam splitter (BS) and is sent to a linearly and a circularly sensitive balanced polarimeter for full polarization analysis; 7) analog-to-digital conversion electronics and data processing. }}
    \label{fig:setup}
\end{figure}

\begin{figure}[ht!]
    \includegraphics[width=0.9\linewidth]{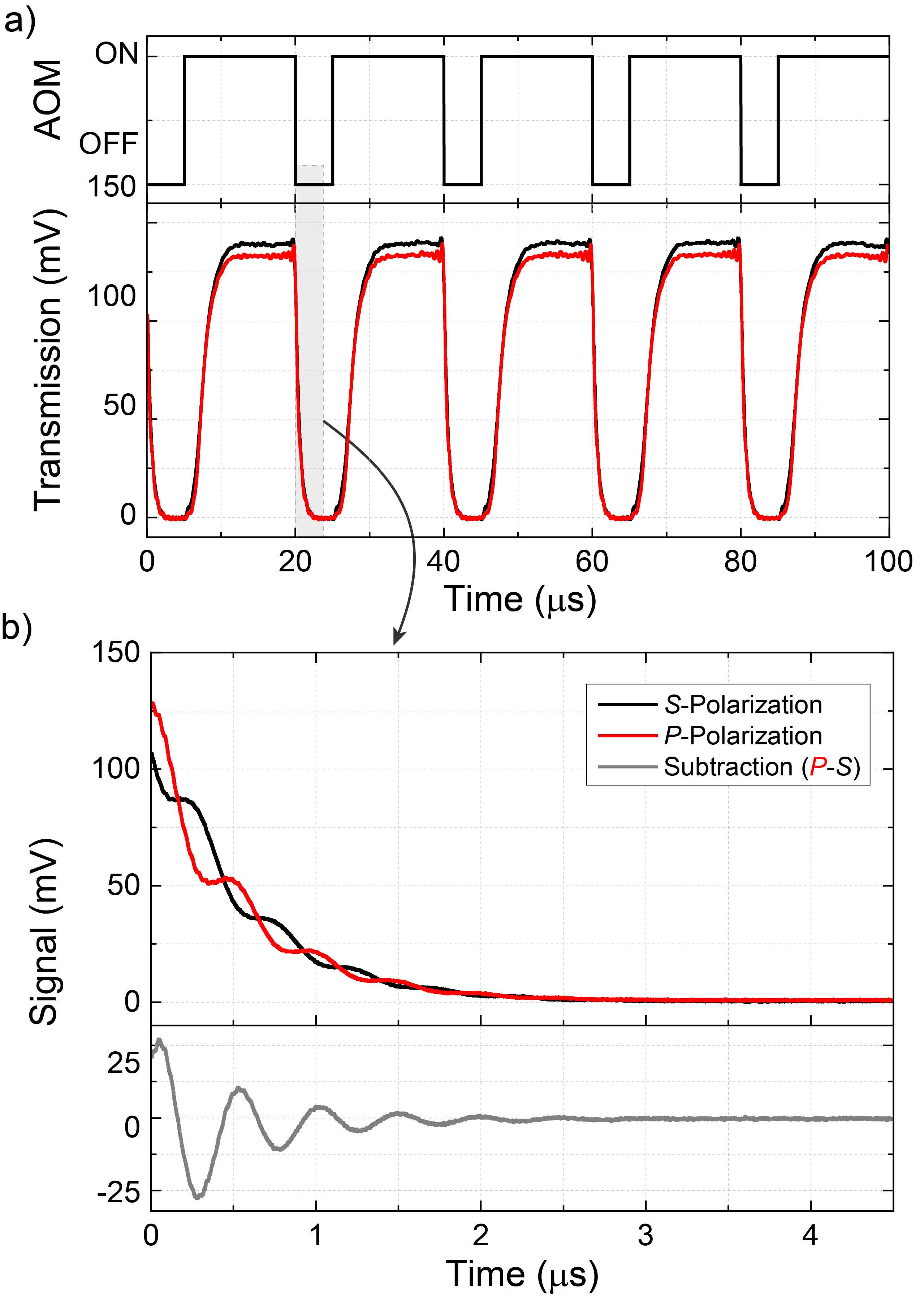}
    \caption{\small{(a) To initiate a ring-down event, the incident radiation, which is frequency locked to the top of a cavity fringe (see Fig.\,\ref{fig:Split}, upper left panel), is abruptly switched off (at time-scales much faster than the cavity lifetime) with the use of an acousto-optic modulator (AOM). (b) Typical CW-CRDP traces, showing polarization beat frequencies generated via the Faraday effect on a SiO$_2$ substrate, as recorded by the $S$ and $P$ channels of a linear balanced polarimeter (Fig.\,\ref{fig:setup}; see also Fig.\,\ref{fig:Split} for a comparison with the theoretical simulations). The data shown here are the result of signal averaging (approximately 200 traces).}}
    \label{fig:CWCRDP}
\end{figure}

In Fig.\,\ref{fig:setup} we present a schematic diagram of the optical setup we use to demonstrate the experimental principles of the CW-CRDP technique. In this work we focus on studying the non-resonant Faraday effect from gaseous species. For this reason, we chose as our intracavity bias anisotropy a non-reciprocal Faraday optical rotation, i.e. $\theta_{_{\rm{F}}}$, which can be generated with the use of a material that has low enough absorption losses and large enough magneto-optic response, to enable sensitive CW-CRDP measurements. We describe in the next section the different options we consider in this work.\\
\indent The ring-down cavity we use has a total length of 0.60\,m and consists of two highly reflective concave mirrors with radii of curvature of 1\,m (FiveNine Optics; specified reflectivity R$\sim$99.9\% at 408\,nm). The mirrors and intracavity optics are mounted on kinematic mounts. To be able to perform Faraday effect measurements of gaseous samples, we install an intracavity, 40\,cm long, solenoid that generates uniform magnetic fields (with strengths of 40\,G per ampere). The solenoid is driven by a DC power supply, and we switch the magnetic field using a relay circuit controlled with a digital-to-analog device (Labjack, U6). The whole cavity is housed within a custom stainless-steel enclosure, sealed with a Plexiglas lid. Anti-reflection (AR) coated flange viewports (Thorlabs, VPCH42-A) are mounted on the enclosure and allow for laser-light access into the cavity and for collecting the cavity reflection required for frequency locking. A rotary vane pump (Kurt J. Lesker, RV224) is connected to the enclosure and is used to pump down the system for measurements, and we monitor the pressure inside the enclosure using a vacuum gauge. We control the injection of gases into the cavity using a needle valve, which is connected directly to the enclosure.\\
\indent The laser source is an external cavity diode laser with a center wavelength of $\lambda = 408$\,nm (Toptica DL-PRO), and we use a set of lenses to achieve spatial mode-matching of the laser beam to the fundamental transverse cavity mode. Our primary focus in this work is the study of non-resonant Faraday effects and, for this reason, we do not pre-stabilize the frequency of our laser to be resonant with a specific transition frequency. However, to establish a CW-CRDP protocol we lock the frequency of the laser to the cavity resonance using a Pound-Drever-Hall (PDH) scheme\,\cite{Black2001}. To prepare the polarization state of the input beam, and to collect the back-reflected one to a photodetector (Thorlabs, PDA10A-EC) for PDH-locking, we use an optical isolator that has side exit ports equipped with polarizing beam splitters (Thorlabs, IO-5-405-LP) followed by a half-wave plate. We generate the error signal using direct current modulation and a PDH module (Toptica, PDD\,110).  The error signal is then input to an analog PID controller (Toptica FALC 110) to provide a fast feedback to the laser-diode current, and to a second analog PID (SRS, SIM960) that creates a slower feedback that acts on the grating of the ECDL through a piezoactuator. The gain and bandwidths of the PID loops are different for low frequencies ($<$10\,Hz) to prevent accumulation of DC-offsets in the error signals that would otherwise prevent a stable lock. To initiate a ring-down event, we use an acousto-optic modulator (AOM; Gooch and Housego 3200-125) driven with a homebuilt AOM-driver, whose RF output can be controlled though a high-bandwidth switch. We analyze the light transmitted through the cavity using two balanced polarimeters: a linearly and a circularly sensitive one, which consist of a half-wave plate (Thorlabs, WPH10ME-405) followed by a Wollaston prism (Thorlabs, WP10), and a quarter-wave plate (Thorlabs, WPQ10ME-405) followed by a Wollaston prism, respectively. All wave-plates and prism are placed within rotation mounts. In each balanced polarimeter, we use short lenses ($f\!=\!50$\,mm) to focus the emerging radiation on separate silicon amplified photo-detectors (Thorlabs, PDA8A). \\
\indent In Fig.\,\ref{fig:CWCRDP} we show typical experimental CW-CRDP traces. With our optical setup and intracavity optics we can achieve ring-down times in the 0.7-1.5\,$\mu$s range, transmitted optical powers of approximately 20-100\,$\mu$W (we are primarily limited by the laser's output power and the impedance mismatching between the reflectivity of the cavity mirrors and the intracavity losses), and with our feedback system and AOM electronics, we are able to generate CW-CRDP traces at repetition rates as high as 50\,kHz.\\ 

\subsection{Data acquisition and signal analysis}
We record and digitize the ring-down traces (photo-detector signals) using two different acqusition modules: an oscilloscope (LeCroy, Wavesurfer 510), which has a maximum acquisition rate of approximately 200\,Hz, an 8-bit resolution per channel, and permits on-board signal averaging; and a 14-bit digitizer (Teledyne, ADQ14DC-2X-PCIE, dual channel DC-coupled operation; sample rates of 2\,GS/s per channel), which has a maximum acquisition rate of approximately 100\,kHz (mainly limited by the data transfer rate), 14-bit resolution per channel, and permits on-board signal averaging. All collected data are transferred into a personal computer, and are processed by a custom Python program (SciPy\,\footnote{https://www.scipy.org}).\\
\indent In this work we use time-domain analysis, based on nonlinear least-squares regression of the recorded time-traces, and all the data we present are the result of such an analysis. In particular, we record signals [$I(t)$] from both polarization channels (Fig.\,\ref{fig:setup} \& Fig.\,\ref{fig:CWCRDP}), and we fit the resulting CRDP-traces using the following model:
\begin{equation}
I(t)=A\,e^{-t/\tau}\,[\,\sin^2(2\pi f\cdot t + \phi)+B\,] + C,
\end{equation}
where $A$ is the amplitude of the trace, $\tau$ is the ring-down time, $f$ is the polarization beat frequency, $\phi$ represents a global phase offset, parameter $B$ takes into account any reduction in modulation depth, and $C$ is a global signal offset. With our acquisition systems and analysis process we can record and monitor in real-time the signal amplitude, ring-down time, frequency, and phase of both polarization channels independently. The subtraction of the signals, which yields a damped sinusoid, is performed through our computer-analysis software.\\
\indent We wish to emphasize here that for CRDP-based portable spectropolarimetric instrumentations and, especially, experiments studying dynamics\,\cite{Sofikitis2015SDR}, online and fast, acquisition and analysis, are crucial. For the purposes of this work we focus on a post-processing time-domain analysis that allows us to thoroughly investigate the sensitivity limits of CW-CRDP experiments for total integration times ranging from a few ms to a few seconds. However, computational algorithms for time-domain analysis are typically slow (in our case we require 10-100\,ms to fit a single trace) and might not be the appropriate methodology for rapid signal analysis (at time scales similar to a single ring-down event), particularly for spectropolarimetry where precision is the key criterion in choosing the appropriate analysis for CRDP techniques. A detailed investigation of the advantages and disadvantages of different signal analysis techniques, to the best of our knowledge, does not exist in the context of CRDP and we will address this in a follow-up work.

\section{Results}
As a proof-of-principle demonstration of the CW-CRDP technique we choose to study the non-resonant Faraday effect from a gaseous species, particularly butane. CW-CRDP relies on the implementation of a large intracavity bias anisotropy that assists in the measurement of the weak Faraday effect of butane (from studies in the literature we know that the expected Verdet constant of butane at 408\,nm is approximately 30\,nrad\,G$^{-1}$\,cm$^{-1}$\,bar$^{-1}$, at 408\,nm). For this reason, we start by investigating different optical materials that can allow for the generation of a large enough intracavity bias circular birefringence while having as low as possible absorption. We are interested in measurements at 408\,nm where typical magneto-optic crystals are expected to have significant absorption losses ($\gtrsim1$\,cm$^{-1}$). We focus our efforts on two optical elements, a SiO$_2$ substrate and a CeF$_3$ crystal, to investigate their magneto-optic response and the possibility of employing them for gas-phase measurements.

\subsection{SiO$_2$ Faraday effect measurement}
\begin{figure}[ht!]
    \includegraphics[width=0.95\linewidth]{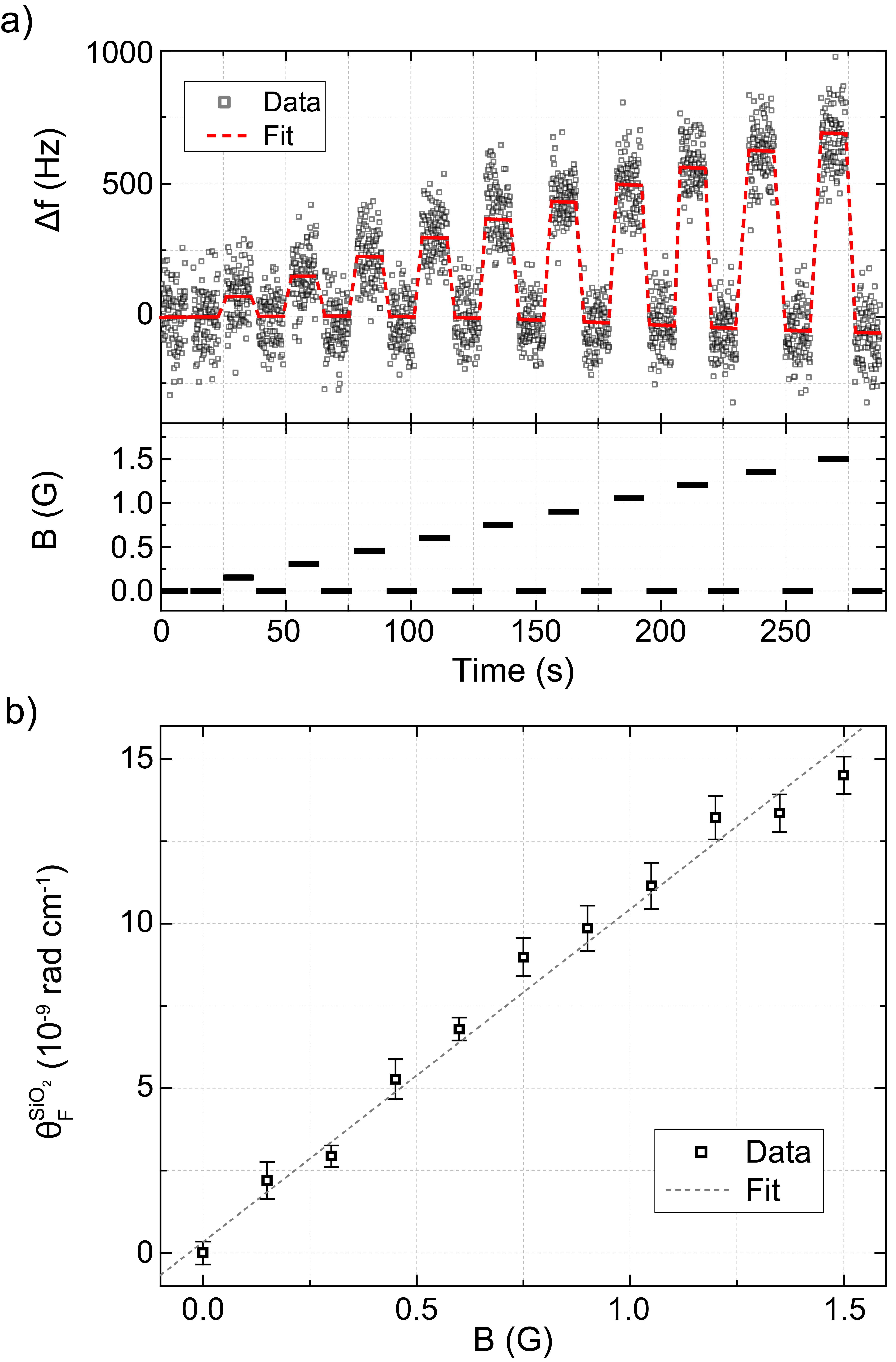}
    \caption{\small{(a) Measurements of frequency shifts [$\Delta \rm{f} = \rm{f}(B)-{f}_0$] with respect to the bias beat frequency induced by a set of permanent magnets placed around a 6.35\,mm thick SiO$_2$ substrate [$\rm{f}_0=2\,121\,123(3)$\,Hz], as a function of an externally applied magnetic field (B). Each measurement point is the result of averaging 5000 CRDP traces, which corresponds to an integration time of 100\,ms per point. (b) Measurements of Faraday optical rotation per unit length for a SiO$_2$ substrate, $\thetaF^{\rm{SiO}_2}$, at 408\,nm, as a function of applied magnetic field. The dashed line is the result of a linear least-squares regression analysis that yields the Verdet constant for SiO$_2$ at 408\,nm: $V^{\rm{SiO}_2}= 10.1(3)\,\mu$rad\,G$^{-1}$\,cm$^{-1}$  (one standard deviation uncertainties).}}
    \label{fig:SIO2data}
\end{figure}

From studies available in the literature we find that the absorption coefficient for SiO$_2$ is estimated to be $<\!10^{-5}$\,cm$^{-1}$\,\cite{Kitamura2007}, while its Verdet constant is $10\,\mu$rad\,G$^{-1}$\,cm$^{-1}$\,\cite{ramaseshan1946determination,ramaseshan1958faraday}, at 408\,nm. Therefore, we expect that high quality SiO$_2$ substrates and AR-coatings can allow for ultra high-finesse cavities, and the Verdet constant is large enough to allow for a sufficiently large $R$-$L$ mode-splitting for applied magnetic fields of approximately a few $\sim$kG (possible with the use of permanent magnets). \\
\indent In Fig.\,\ref{fig:SIO2data} we present measurements of the (non-resonant) Faraday effect of a 6.35(1)\,mm thick, AR-coated SiO$_2$ substrate (FiveNines Optics; AR coated by FiveNine Optics with specified  R$<0.01$\%), which is also the first demonstration of the CW-CRDP technique. To generate a large bias Faraday optical rotation we use permanent magnets (K\&J Magnetics, Inc.) attached directly to the mount holder of the substrate. For measuring precisely the Verdet constant of the substrate we use an additional, homemade, solenoid [with a length of $2.53(1)$\,cm, and a diameter of $3.05(1)$\,cm], which we place around the SiO$_2$ substrate. We calibrate the magnetic field of the solenoid using a Hall probe magnetometer (Hirst Magnetic Instruments GM05), and use a USB controlled digital-to-analog device (Labjack, U6) to drive the solenoid and scan the magnetic field it generates. The permanent magnets allows us to generate a large bias Faraday optical rotation, yielding a bias beat frequency of $\rm{f}_0=2\,121\,123(3)$\,Hz, and using the smaller solenoid we are able to induce frequency shifts around this central beat frequency [Fig.\,\ref{fig:SIO2data}\,(a)]. For each magnetic field strength we collect 100 measurement points, each of which is the result of averaging 5000 CRDP traces corresponding to an integration time of 100\,ms per measurement point. Using these, we obtain the resulting frequency shift [$\Delta \rm{f}={\rm{f}}(B)- {\rm{f}}_0$] as a function of magnetic field strength, which in turn yields the Faraday optical rotation for SiO$_2$ [Fig.\,\ref{fig:SIO2data}\,(b)] that is directly proportional to the Verdet constant of the SiO$_2$ substrate. Importantly, we observe in our measurements frequency drifts that can be associated with mechanical drifts, but our subtraction procedure removes these and allows us to observe a linear dependence of the measured optical rotation on the applied magnetic field. With this procedure, we measure the Verdet constant to be $V^{\rm{SiO}_2}=10.1(3)\,\mu$rad\,G$^{-1}$\,cm$^{-1}$, at 408\,nm, which is in accord with results available in the literature (Refs.\,\cite{ramaseshan1946determination,ramaseshan1958faraday}). The measurement error bars are derived through propagation of the errors ascribed to the measured polarization beat frequencies.

\subsection{CeF$_3$ Faraday effect measurement}
\begin{figure}[ht!]
    \includegraphics[width=\linewidth]{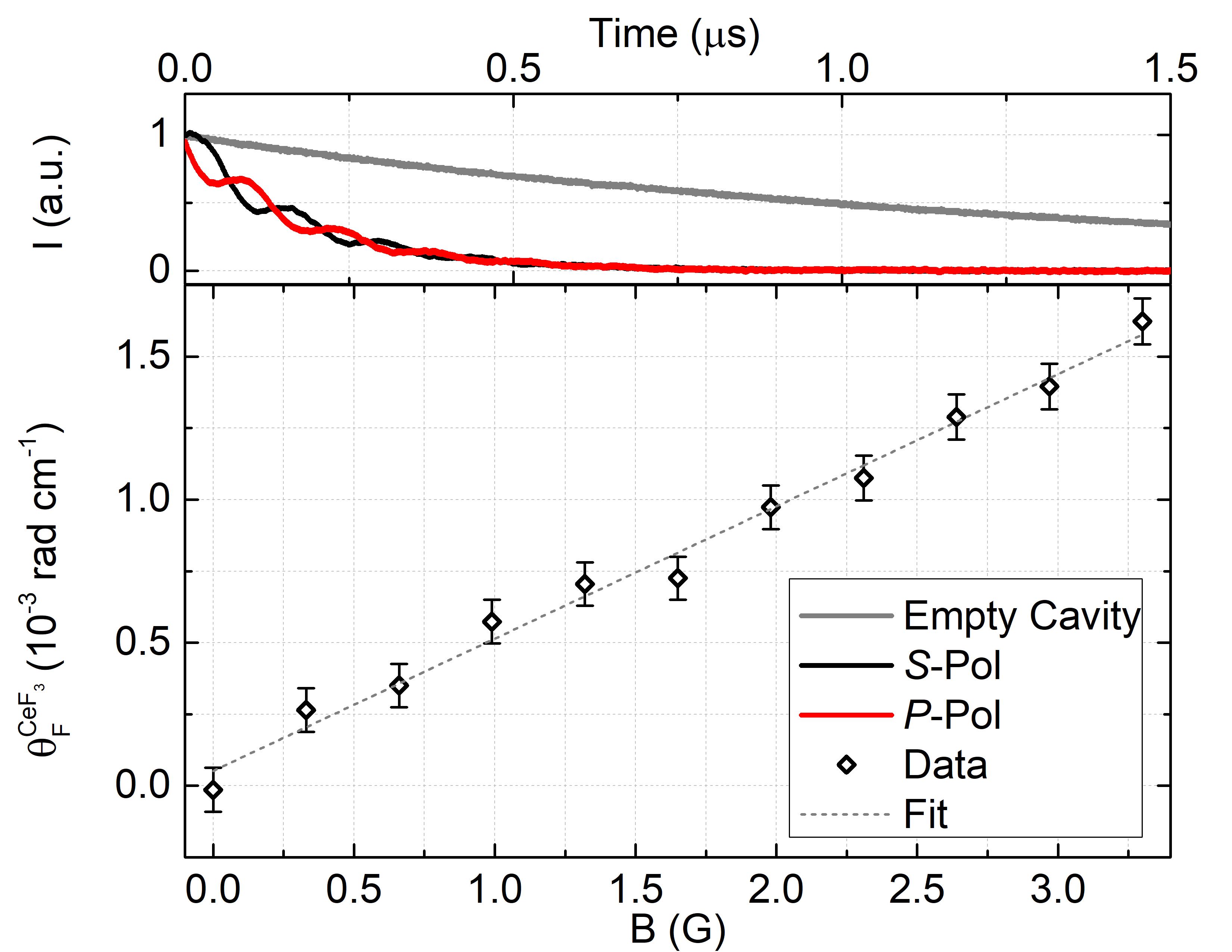}
    \caption{\small{\textit{Upper panel -} CW-CRDP traces for an empty cavity vs. one with an intracavity 1.20\,mm thick AR-coated CeF$_3$ crystal. The bias polarization beat frequency, $\rm{f}_0\!=\!3\,575\,614(115)$\,Hz, is the result of large Faraday optical rotation generated using a set of permanent magnets. \textit{Lower panel -} Measurements of Faraday optical rotation per unit length for the CeF$_3$ crystal, $\theta_{\rm{F}}^{\rm{CeF}_3}$, at 408\,nm, as a function of applied magnetic field. The dashed line is the result of a linear least-squares regression analysis that yields the Verdet constant for CeF$_3$ at 408\,nm: $V^{\rm{CeF}_3}= 462(16)\,\mu$rad\,G$^{-1}$\,cm$^{-1}$  (one standard deviation uncertainties).}}
    \label{fig:CeF3data}
\end{figure}

Garnet single crystals, such as yttrium-iron-garnet (YIG) and terbium-gallium-garnet (TGG), are magneto-optic materials typically used in Faraday rotators, due to their transparency in the visible and NIR optical ranges and their high Verdet-constant values. However, all these materials exhibit significant losses at wavelengths $\!<\!600$\,nm. Recent studies have reported on alternative materials suitable for Faraday rotators at near-UV optical wavelengths, such as CeF$_3$ and PrF$_3$\,\cite{SHIMAMURA2004208,Molina2011}. For CeF$_3$, in particular, the absorption coefficient at 408\,nm is predicted to be $\lesssim\!10^{-1}$\,cm$^{-1}$, and its Verdet constant is predicted to be approximately 390\,$\mu$rad\,G$^{-1}$\,cm$^{-1}$\,\cite{Molina2011}. To verify these numbers and, therefore, investigate whether we can employ this crystal for inducing an intracavity bias anisotropy for CW-CRDP at 408\,nm (and similar wavelengths), we use a 1.20(1)\,mm thick, AR-coated CeF$_3$ crystal (E-Crystal Co., Ltd.; AR-coated by E-Crystal Co., Ltd. with specified  R$<0.2$\%).\\
\indent In Fig.\,\ref{fig:CeF3data} we present measurements of the (non-resonant) Faraday effect of crystalline CeF$_3$ as a function of magnetic field. Using the crystal, we observe ring-down times of approximately $\tau_{_{{\rm{CeF}}_3}}\!\approx\!200$\,ns (compared to the attainable ring-down time of $\tau\!\approx\!1.5\,\mu$s for the empty cavity), which suggests overall losses of $\approx\!0.14$\,cm$^{-1}$ at 408\,nm. Although, these losses are the combined result of scattering losses from the AR-coatings and of material absorption, we attribute these to absorption. To measure precisely the material's Verdet constant we follow a similar measurement approach as described above. A set of permanent magnets allows us to generate a large bias Faraday optical rotation frequency, yielding a bias beat frequency of $\rm{f}_0=3\,575\,614(115)$\,Hz, and using the smaller solenoid we are able to induce frequency shifts around this central bias beating frequency. We measure the Verdet constant to be $V^{\rm{CeF}_3}=462(16)\,\mu$rad\,G$^{-1}$\,cm$^{-1}$, at 408\,nm. Our findings are in relative agreement with the results presented in Ref.\,\cite{Molina2011}.  

\subsection{Gaseous butane Faraday effect measurement}
\begin{figure}[ht!]
    \includegraphics[width=0.95\linewidth]{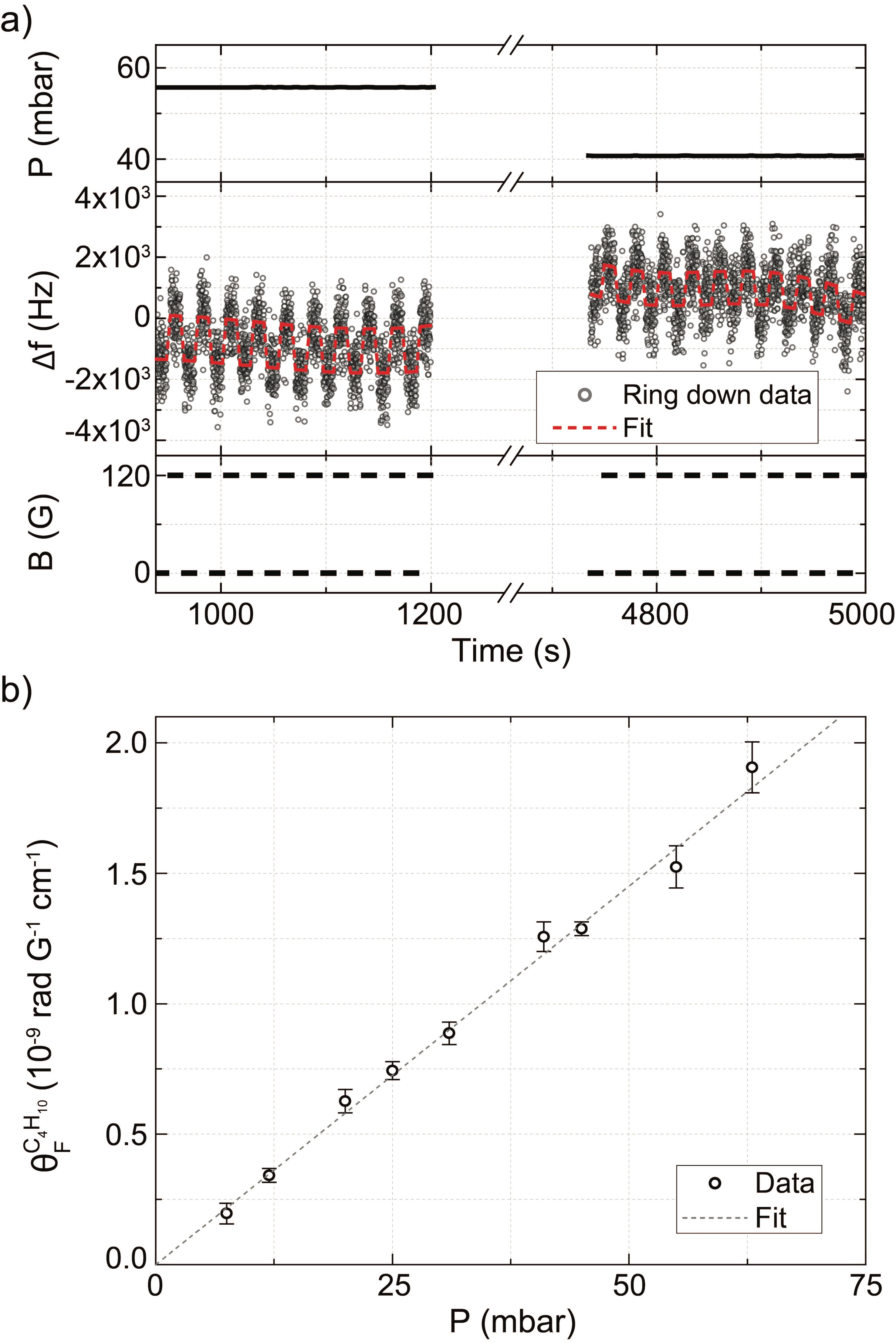}
    \caption{\small{(a) Measurements of frequency shifts [$\Delta \rm{f} = \rm{f}(B)-{f}_0$] with respect to the bias beat frequency induced by the Faraday rotation from a SiO$_2$ substrate [$\rm{f}_0\!=\!2\,858\,570(20)$\,Hz], due to the non-resonant Faraday effect of gaseous butane at 408\,nm, for two different gas pressures [55.6(1)\,mbar and 40.7(1)\,mbar] and for an applied magnetic field of 120\,G. Each measurement point is the result of averaging 5000 CRDP traces, which corresponds to an integration time of 100\,ms per point. (b) Measurements of Faraday optical rotation per unit length for gaseous butane, $\theta_{\rm{F}}^{\rm{C}_4\rm{H}_{10}}$, at 408\,nm, as a function of pressure for a magnetic field of 120\,G. The dashed line is the result of a linear least-squares regression analysis that yields the Verdet constant for butane at 408\,nm: $V^{\rm{C}_4\rm{H}_{10}}= 29.0(7)$\,nrad\,G$^{-1}$\,cm$^{-1}$\,bar$^{-1}$  (one standard deviation uncertainties). }}
    \label{fig:butaneResults}
\end{figure}
For the measurement of the Faraday effect in gaseous butane we induce an intracavity bias anisotropy with the SiO$_2$ substrate (Fig.\,\ref{fig:SIO2data}). We use a similar measurement procedure as described before and employ a homemade solenoid (40\,cm) for the generation of adjustable magnetic fields (Fig.\,\ref{fig:setup}). We use an external power supply to control the current of the solenoid and a relay circuit to controllably switch the magnetic field on and off. \\
\indent In Fig.\,\ref{fig:butaneResults} we present measurements of the (non-resonant) Faraday effect of butane (C$_4$H$_10$) as a function of pressure. For each pressure, we perform approximately 10 magnetic field cycles, and for each magnetic field cycle we collect 100 measurement points, each of which is the result of averaging 5000 CRDP traces (corresponding to an integration time of 100\,ms per measurement point). Using these, we obtain the frequency difference with respect to the bias Faraday beat frequency generated using a SiO$_2$ substrate [$\Delta \rm{f}\!=\!\rm{f(B)}-f_0$, with $\rm{f}_0=2\,858\,570(20)$\,Hz]. This process ensures that we remove any contribution to the signal originating from the permanent magnets used to create this large bias Faraday anisotropy. We verify the linear dependence between the observed Faraday optical rotation (per unit length) and butane's vapor pressure, and we measure its Verdet  constant to be $V^{\rm{C}_4\rm{H}_{10}}= 29.0(7)$\,nrad\,G$^{-1}$\,cm$^{-1}$\,bar$^{-1}$, at 408\,nm and 21\,$^\circ$C [Fig.\,\ref{fig:butaneResults}\,(b)]. Our measurements are in accord with results available in the literature (Refs.\,\cite{ingersoll1954faraday,Ingersoll1956}).

\section{Sensitivity} 
\subsection{Cram\'er-Rao Lower Bound: fundamental sensitivity limit in CRDP}
\begin{figure}[ht!]
    \includegraphics[width=\linewidth]{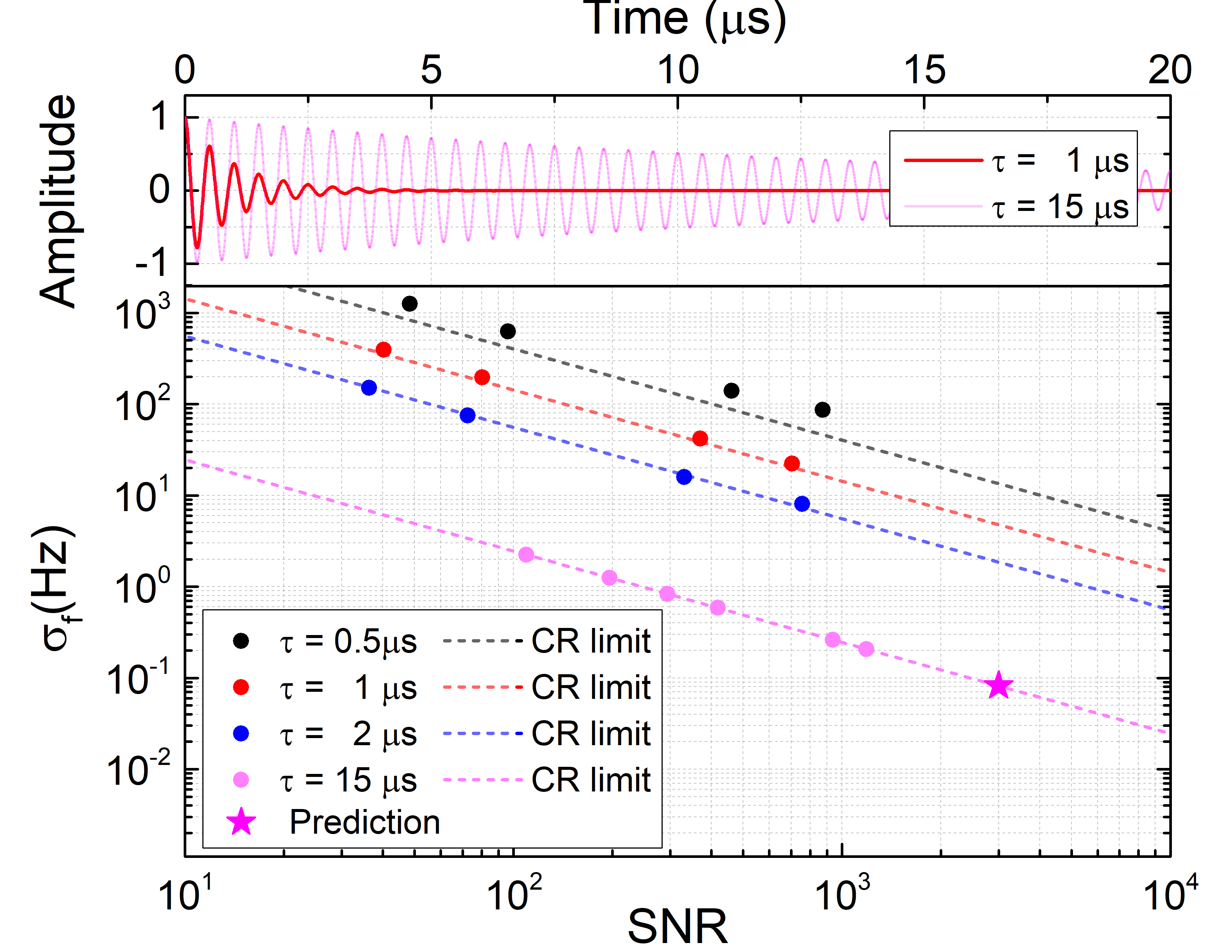}
    \caption{\small{\textit{Upper panel -} Simulated CRDP traces generated using an arbitrary waveform generator (8-bit resolution) for different ring-down times and a fixed beat frequency of 2\,MHz, which are recorded using a 14-bit digitizer. \textit{Lower panel -} Frequency estimation errors obtained using a time-domain analysis of the simulated CRDP traces. The dashed lines are the estimation limits set by the Cram\'er-Rao (CR) lower bound (Eq.\,\ref{eq:CRLB}) for the specified decay times ($\tau$) and signal-to-noise ratios (SNR) of the simulated signals. The star-shaped point labeled as ``Prediction", indicates that in CRDP experiments with attainable ring-down times of $\tau\!=\!15\,\mu$s and signal acquisitions with SNR=3000, one can achieve sub-100\,mHz frequency sensitivities.	}}
    \label{fig:CRLB}
\end{figure}

The noise in any CRDP-based polarimetric measurement will contain contributions from both technical and intrinsic (fundamental) sources. For typical experimental setups, similar to the one we use for the measurements, technical noise sources are predominantly of mechanical and acoustic nature with non-white power noise spectral densities, and their associated frequency/phase drifts will effectively be the major limiting factor influencing high-sensitivity measurements. However, we wish to consider here the fundamental sensitivity of any CRDP scheme that is directly related to the measurement sensitivity of the polarization beat frequency. While we generally focus on time-domain analysis of the CRDP signals, nothing precludes us from using alternative analysis schemes, such as frequency-domain approaches. Notwithstanding, independent of which analysis methodology we choose, the fundamental limit for the statistical uncertainty of determining the beat frequency from a given CRDP trace will be given by the Cram\'er-Rao lower bound (CRLB)\,\cite{Yao1995}, which sets the lower limit on the variance $\sigma^2_{\rm{f}}$ of any frequency estimator. The CRLB condition for the frequency extracted from a discrete damped sinusoid - the CW-CRDP signal in our case as obtained from the balanced detection stage - is given by\,\cite{Yao1995,Gemmel2010,Koch2015,Hunter2018},
\begin{equation}
    \sigma^2_{\rm{f}} = \dfrac{6}{(2\pi)^2\,{\rm{SNR}}^2\,{\rm{f}}_{_{\rm{BW}}}\,{\rm{T}}_{\rm{meas}}^3} \chi\left(\tau/{\rm{T}}_{\rm{meas}}\right),
    \label{eq:CRLB}
\end{equation}
where SNR is the measured signal-to-noise ratio, which is defined as the ratio between the signal amplitude and the standard deviation of the noise (effectively the electronic noise of the acquisition system); $\rm{f}_{_{\rm{BW}}}$ is the sampling-rate-limited bandwidth of the measurement; ${\rm{T}}_{\rm{meas}}$ is the measurement time window; and $\chi(r)$ is a corrective factor that takes into account the signal decay that is given by,
\begin{equation}
    \chi(r) = \dfrac{e^{2/r} - 1}{3r^3\,\cosh(2/r)-3r\left(r^2+2\right)}.
\end{equation}
The factor $\chi\left(\tau/{\rm{T}}_{\rm{meas}}\right)$ in Eq.\,\ref{eq:CRLB} is a compensation factor that penalizes measurement of the tails of the exponential decay when the signal has effectively died out. Importantly, Eq.\,\ref{eq:CRLB} assumes that the period of the oscillation is much shorter than the decay constant of the envelope and that a sufficient number of oscillations occurs in the signal, without limiting the estimation accuracy. We emphasize here that Eq.\,\ref{eq:CRLB} dictates that any noise sources affecting the experimental measurements are contributing to the fundamental CRLB limit through their effect on the SNR of the CRDP signal.\\
\indent To test the validity of Eq.\,\ref{eq:CRLB} as the appropriate estimator of the fundamental sensitivity for a frequency-based CRDP measurement, we use an arbitrary waveform generator (AWG; Tektronix AWG7122C, 10-bit resolution) to simulate CRDP signals and determine the sensitivity in measuring their beat frequency. In particular, we simulate CRDP signals with a constant beat frequency of 2\,MHz and variable signal-to-noise ratios (SNR) and ring-down times. We use the 14-bit digitizer (Teledyne, ADQ14DC-2X-PCIE) to record (over a fixed time window of approximately $\sim\!7\tau$) and integrate AWG-generated CRDP traces, which are analyzed using a time-domain approach to obtain their amplitude, frequency and respective SNR. Although we vary the SNR of the simulated signals by the introduction of white gaussian noise through the AWG, we use the time domain analysis for the estimation of the SNR value we use in Eq.\,\ref{eq:CRLB}. \\
\indent In Fig.\,\ref{fig:CRLB} we show two special cases of the simulated CRDP traces together with the results of our analysis. We confirm that the frequency estimation errors coincide with the CRLB limit (Eq.\,\ref{eq:CRLB}) for a wide range of ring-down decay times and SNRs (for ${\rm{f}}_{_{\rm{BW}}}=500$\,MHz). Most importantly, we simulate a CRDP trace with $\tau\!=\!15\,\mu$s, and confirm that the CRLB limit for a CRDP frequency measurement of such a trace with an SNR=3000 to be at the sub-Hz levels, corresponding to sub-ppm fractional uncertainties (i.e. a $\sim\!100$\,mHz uncertainty on a measurement of a 2\,MHz beating frequency). We also verify that for CRDP signals with short ring-down times, e.g.\,\,0.5\,$\mu$s, where one observes only a limited number of oscillations within the whole acquisition window of the ring-down signal, the results start to deviate from the CRLB limit. \\
\indent Equation\,\ref{eq:CRLB} sets the frequency detection limit and, therefore, the CRLB-related optical rotation sensitivity limit, $\sigma_{\theta}$, is equal to,
\begin{equation}
    \sigma_{\theta} = \frac{\pi}{\rm{FSR}}\sigma_{\rm{f}}.
\end{equation}
Thus, the CRLB limit can used as a guide for designing CW-CRDP experiments (i.e. choosing the appropriate finesse, the strength of intracavity anisotropy, the electronics and acquisition system), for a particular polarimetric application. 


\subsection{CW-CRDP experimental sensitivity limits}
\begin{figure}[ht!]
    \includegraphics[width=\linewidth]{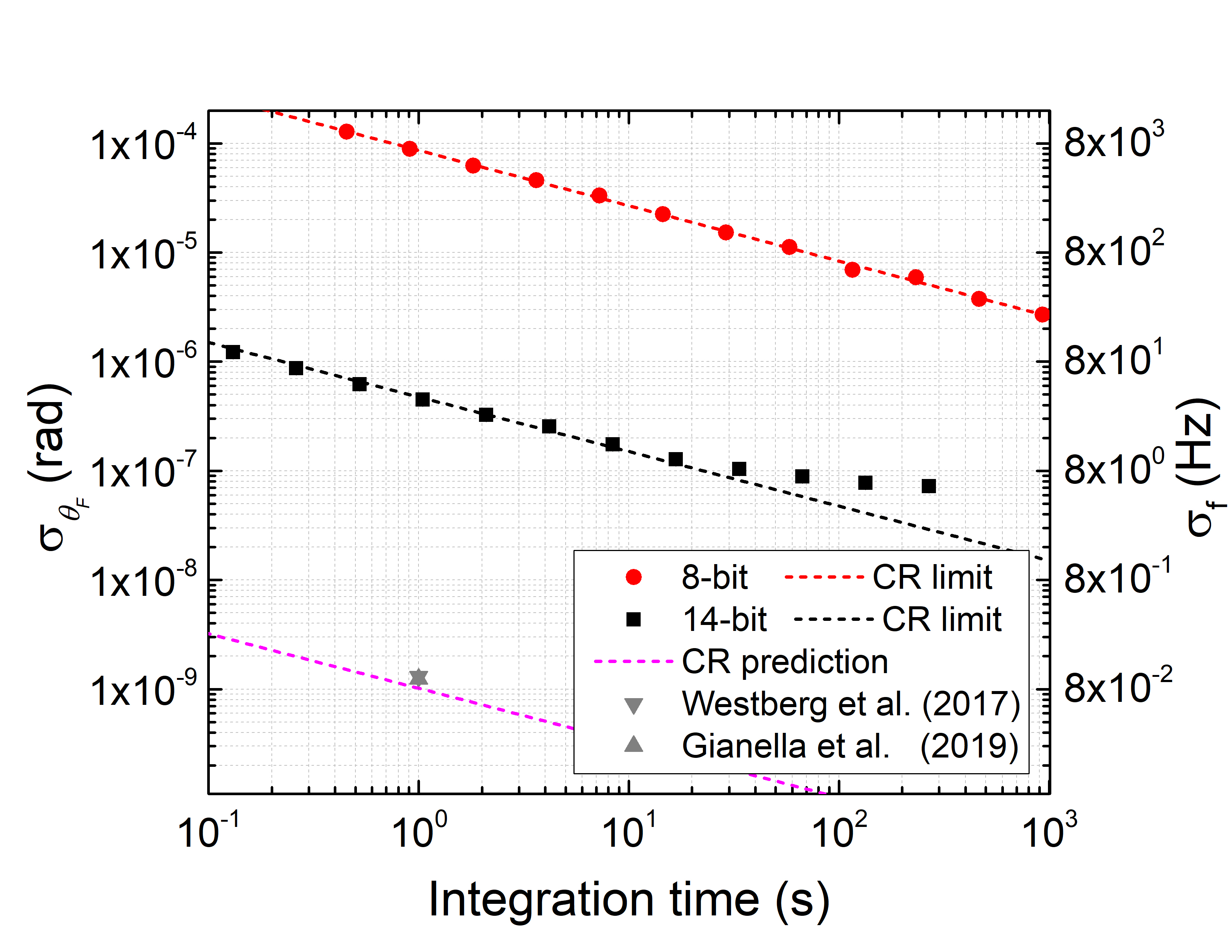}
    \caption{\small{Allan variance of CW-CRDP polarization beat frequency measurements, $\sigma^2_{\rm{f}}$, of the non-resonant Faraday effect of an SiO$_2$ substrate, using two different acquisition modalities (of 8-bit vs.\,14-bit resolution). At short integration times white noise dominates and the variance decreases proportionally as the inverse of the square-root of the integration time. Mechanical drifts influence the stability of the measurements beyond an optimum integration time of $\approx30$\,s. The dashed lines (black, red, magenta) are the Cram\'er-Rao limits for our experimental parameters and for a predicted set of experimental parameters. 
    }}
    \label{fig:AllanW}
\end{figure}

\indent We proceed by analysing the sensitivity and stability of the experimental apparatus using an Allan-variance methodology, which allows us to study the temporal characteristics of the frequency measurements. In Fig.\,\ref{fig:AllanW} we present an Allan plot of CW-CRDP measurements of the polarization beat frequency induced by the Faraday effect in the SiO$_2$ substrate under the influence of a constant magnetic field (induced by a set of permanent magnets). We use two separate acquisition systems of different resolution and sampling rates to acquire and integrate CW-CRDP signals for over $\sim$1000\,s, which enables us to examine the stability of our apparatus and the validity of the CRLB limit for the different acquisition conditions. Using time-domain analysis (Sec.\,III\,B), we estimate the uncertainty in our frequency measurements over a specific integration time. Using the high (14-bit) resolution acquisition system, we observe that our system exhibits a white-noise like behaviour with a slope of approximately $35\,\rm{Hz}/\sqrt{\rm{Hz}}$, which corresponds to polarimetric sensitivities of approximately $448\times10^{-9}\,\rm{rad}/\sqrt{\rm{Hz}}$ per round-trip (the cavity's finesse is $\approx$1200), and this behaviour holds for integration times of up to $\sim$30\,s, where drifts associated with our frequency stabilization system do not allow for further sensitivity improvements. Using the low (8-bit) resolution acquisition system we are unable to observe these drifts that fundamentally limit our measurements due to the low signal resolution, and, thus, sensitivity. However, for both acquisition conditions, we observe that our experimental results are in agreement with Eq.\,\ref{eq:CRLB} (particularly, using the high resolution acquisition system we achieve a SNR$\approx\!700$ within 1\,s of integration time for repetition rates of 50\,kHz). We note here that while our sampling bandwidth is 500\,MHz (i.e. ${\rm{f}}_{_{\rm{BW}}}$ in Eq.\,\ref{eq:CRLB}), our photodetectors have a bandwidth of 50\,MHz, and, therefore, consecutive sampling points are not statistically independent. However, this affects the attainable SNR from the time-domain analysis\,\cite{Huang2013}, as supported by our agreement with the CRLB prediction. As a comparison, we note here that pulsed-CRDP techniques have achieved polarimetric sensitivities at the $\sim10^{-6}$\,rad levels for several minutes of integration time using optical cavities of similar finesse as the one we use here\,\cite{Muller2002,Sofikitis2014}, clearly demonstrating the benefits of employing CW-laser sources. In addition, using our experimental conditions, we predict the photon shot-noise limit for our CRD-based polarimetric measurements to be $\sim\!4\times10^{-9}\,\rm{rad}/\sqrt{\rm{Hz}}$ per round-trip (for these measurements we collect approximately $\sim$20\,$\mu$W of optical power, and estimate this limit using the total number of photons available from all ring-down events within 1\,s)\,\cite{Huang2013}. \\
\indent To demonstrate the merits of CW-CRDP, in Fig.\,\ref{fig:AllanW} we include for comparison the reported polarimetric sensitivities of two recent state-of-the-art PD-CRDS-related works. In particular, Westberg and Wysocki\,\cite{Westberg2017} using an optical cavity with $\mathcal{F}\!\approx\!50000$ demonstrated Faraday optical rotation sensitivities of $\sim1.3\times10^{-9}$rad/$\sqrt{\rm{Hz}}$ (per cavity round-trip, for 1\,s of integration time), at 762\,nm. Similarly, Gianella et al.\,\cite{Gianella2019} using an optical cavity with $\mathcal{F}\!\approx\!175000$, demonstrated almost identical Faraday optical rotation sensitivities of $\sim1.24\times10^{-9}$\,rad/$\sqrt{\rm{Hz}}$ (per cavity round-trip, for 1\,s of integration time), at 1506\,nm. Using our analysis on the fundamental frequency sensitivity limits of CRDP (``Prediction" point in Fig.\,\ref{fig:CRLB}), we predict that a CW-CRDP experiment realized using a 0.60\,m long two-mirror optical cavity with $\mathcal{F}\!\approx\!23560$, resulting in ring-down times of 15\,$\mu$s, and which yields signals with an SNR of 3000 within 1\,s of integration time (we assume here only white noise), one can achieve Faraday optical rotation sensitivities of $\sim\!1\times10^{-9}$\,rad/$\sqrt{\rm{Hz}}$ (we assume here a sampling-rate-limited bandwidth of 500\,MHz). Therefore, one can achieve similar state-of-the-art sensitivities with a CW-CRDP polarimetric approach using cavities of significantly lower finesse, opening the possibility for highly sensitive polarimetry to spectral regimes where high quality optics might not be available and/or material losses can be higher. Finally, it is important to emphasize that for our prediction, where we consider a reasonably attainable SNR for 1\,s of integration time, we do not consider the case of correlated noise cancellations as a result of the balanced detection scheme\,\cite{Westberg2017,Gianella2019} or possible SNR improvements that arise from the application of rapid signal reversals\,\cite{Sofikitis2014,Bougas2015,Hayden2018}. Moreover, the experimental parameters we choose for our prediction can be realized by several improvements in our current experimental setup (by improving the cavity finesse, increasing the transmitted optical power, using low-noise photodetectors, and improving our frequency-locking system), but these correspond to one set of possible parameters that can be selected to optimize the CRLB (Eq.\,\ref{eq:CRLB}) and yield state-of-the-art polarimetric sensitivities.

\subsection{Alternative modalities for CW-CRDP}
\indent In the theory section (Sec.\,II) we discuss the principles of CW-CRDP for the case of a CW laser source with a linewidth that is significantly smaller than that of the optical cavity and of an intracavity anisotropy that results in a substantial overlap between the $R$-$L$ cavity modes. Under these conditions we show how the incident radiation can couple into and build-up within that cavity by exciting only one particular cavity polarization eigenmode at a time, and how this in turn defines the amplitude, phase, and modulation depth of the resulting CW-CRDP signals. From Fig.\,\ref{fig:Split} we see that these parameters can change significantly as a function of laser detuning with respect to the $R$-$L$ mode splitting, and for this reason a robust frequency stabilization scheme is required to ensure minimal frequency deviations that result in signal phase and signal fluctuations, which in turn hinder the sensitivity of a CW-CRDP measurement. Furthermore, the signal parameters also depend on the strength of the intracavity anisotropy (e.g. $\thetaF$) and on the cavity linewidth; if the strength of the intracavity anisotropy is large enough to result in mode splittings larger than a few cavity linewidths (or, equivalently, if the cavity finesse is high enough to result in a narrow cavity linewidth), substantial mode overlap is not possible and, thus, CW-CRDP measurements are unrealistic using a narrow-linewidth laser source. \\
\indent A possible solution towards highly sensitive CW-CRDP instrumentation that does not require complicated electronics and resolves the above-mentioned issues, is to use a single-frequency laser with a linewidth that is significantly broader than that of the cavity $R$-$L$ mode-splitting. In this case, the incident radiation excites both modes coherently and the frequency-stabilization scheme can be replaced by an intensity threshold comparator-circuit that switches off the laser radiation once a transmission threshold is recorded, similarly to what has been done in CW-CRDS demonstrations\,\cite{ROMANINI1997}. For example, for optical cavities as the one we present here and similar polarization mode-splittings of the order of 1-2\,MHz, the laser light linewidth should be at least an order of magnitude larger (i.e.\,10-20\,MHz). Most importantly, using this approach one is not anymore limited by the finesse of the cavity and the desired mode-splitting can be significantly larger than the cavity linewidth, contrary to the case of a CW laser source with a narrow linewidth for which one must generate a mode-splitting of the order of the cavity linewidth for optimal mode coupling. Free-running diode and distributed-feedback lasers typically have linewidths of $<$5\,MHz, while the design approach for commercially available external cavity diode lasers is to reduce the diode-laser linewidth as much as possible (with current linewidths at the sub-100\,kHz range). A direct method to achieve the desired broad linewidth is to broaden the linewidth of the laser source in a controllable fashion by, e.g., modulating the laser current using white noise with high bandwidth. This method is, in principle,  applicable with both diode and distributed-feedback lasers (however, diode-based lasers might face the problem of mode fragmentation due to the noise modulation). There exist several additional modulation approaches that are suitable for CW-CRDP, such as the case of modulating the laser current at the frequency of the mode-splitting, and we will explore these in future works. However, we consider the approach of broadening the linewidth of the laser source to be larger than the mode-splitting, as the most suitable one for portable spectropolarimetric instrumentations.\\ 
\indent An alternative option is to implement an optical-feedback technique, whereby the circulating wave within the cavity is allowed to return to the laser source to injection-seed it, inducing, thus, a direct frequency-locking without the need for the generation of an error signal and the associated feedback electronics. This is the principle behind optical feedback cavity-enhanced absorption spectroscopy (OF-CEAS)\,\cite{Morville2005}, which has already been employed in CRD-based spectroscopic\,\cite{Burkart2014} and polarimetric\,\cite{Gianella2017} measurements. In this case, the linewidth and central frequency of the laser source are effectively matched and (phase-)stabilised to that of the cavity, respectively, allowing for optimal injection and the reduction of phase-fluctuations related to frequency drifts of either the laser or the cavity.\\

\section{Conclusions}
In this paper, we present the general principles of CW-CRDP, a CRD-based polarimetric scheme that employs CW laser sources and benefits from the use of large intracavity polarization anisotropies for highly sensitive polarimetric measurements. Contrary to PD-CRDS polarimetric approaches that are effectively intensity-based measurement schemes (since these monitor changes in the ring-down decay time), CRDP schemes are frequency-based measurements where sensitive detection of shifts in their induced polarization beat frequency enables highly sensitive polarimetric measurements. CW-CRDP builds upon the measurement methodologies that have been developed within the context of chiral pulsed-CRDP\,\cite{Mueller2000,Sofikitis2014,Bougas2015}, and we show how the use of CW laser sources allows for significant gains in signal intensity and data acquisition rates, enabling shot-noise limited measurements that are not easily attainable using pulsed laser sources. Furthermore, we discuss in depth the fundamental limits of CRDP protocols, and demonstrate how CW-CRDP instruments can perform highly sensitive measurements even with the use of optical cavities with modest finesse. We confirm the principles and merits of CW-CRDP by measuring non-resonant Faraday optical rotations from solid and gaseous samples using a prototype design.\\
\indent In conclusion, CW-CRDP is a powerful and versatile technique suitable for the measurement of reciprocal and non-reciprocal birefringence and dichroism, linear and circular. We believe that CW-CRDP is the ideal modality for portable spectropolarimetric instrumentations as it allows for time-resolved, highly sensitive, and cost-effective operation at a broad spectral region. Future work will focus on improvements in system design and sensitivity, and on the possibility of coupling CW-CRDP with chromatographic techniques, towards the development of a portable instrument suitable for breath analysis and monitoring in clinical settings\,\cite{Wang2015,Henderson2018}, and for trace gas analysis and monitoring of paramagnetic species in field settings\,\cite{Hayden2018}.

\section{Acknowledgments}
\indent This work was supported by the European Commission Horizon 2020, ULTRA-CHIRAL project (grant no. FETOPEN-737071), and by the European Union's Seventh Framework Programme for Research, Technological Development, and Demonstration, under the project ERA.Net RUS Plus, grant agreement no.189 (EPOCHSE). L.B. is grateful to E. G. V\'{\i}llora and her colleagues for the help and support in obtaining the CeF$_3$ crystal, and specially thanks Geoffrey Iwata and Kevin Morby for insightful discussions.

\section*{Data Availability Statement}
The data that support the findings of this study are available from the corresponding author upon reasonable request.

\bibliography{20200326LBCWCRDParxivSubmission} 

\begin{thebibliography}{63}%
\makeatletter
\providecommand \@ifxundefined [1]{%
 \@ifx{#1\undefined}
}%
\providecommand \@ifnum [1]{%
 \ifnum #1\expandafter \@firstoftwo
 \else \expandafter \@secondoftwo
 \fi
}%
\providecommand \@ifx [1]{%
 \ifx #1\expandafter \@firstoftwo
 \else \expandafter \@secondoftwo
 \fi
}%
\providecommand \natexlab [1]{#1}%
\providecommand \enquote  [1]{``#1''}%
\providecommand \bibnamefont  [1]{#1}%
\providecommand \bibfnamefont [1]{#1}%
\providecommand \citenamefont [1]{#1}%
\providecommand \href@noop [0]{\@secondoftwo}%
\providecommand \href [0]{\begingroup \@sanitize@url \@href}%
\providecommand \@href[1]{\@@startlink{#1}\@@href}%
\providecommand \@@href[1]{\endgroup#1\@@endlink}%
\providecommand \@sanitize@url [0]{\catcode `\\12\catcode `\$12\catcode
  `\&12\catcode `\#12\catcode `\^12\catcode `\_12\catcode `\%12\relax}%
\providecommand \@@startlink[1]{}%
\providecommand \@@endlink[0]{}%
\providecommand \url  [0]{\begingroup\@sanitize@url \@url }%
\providecommand \@url [1]{\endgroup\@href {#1}{\urlprefix }}%
\providecommand \urlprefix  [0]{URL }%
\providecommand \Eprint [0]{\href }%
\providecommand \doibase [0]{http://dx.doi.org/}%
\providecommand \selectlanguage [0]{\@gobble}%
\providecommand \bibinfo  [0]{\@secondoftwo}%
\providecommand \bibfield  [0]{\@secondoftwo}%
\providecommand \translation [1]{[#1]}%
\providecommand \BibitemOpen [0]{}%
\providecommand \bibitemStop [0]{}%
\providecommand \bibitemNoStop [0]{.\EOS\space}%
\providecommand \EOS [0]{\spacefactor3000\relax}%
\providecommand \BibitemShut  [1]{\csname bibitem#1\endcsname}%
\let\auto@bib@innerbib\@empty
\bibitem [{\citenamefont {Eliel}\ and\ \citenamefont
  {Wilen}(2008)}]{Eliel2008}%
  \BibitemOpen
  \bibfield  {author} {\bibinfo {author} {\bibfnamefont {E.~L.}\ \bibnamefont
  {Eliel}}\ and\ \bibinfo {author} {\bibfnamefont {S.}~\bibnamefont {Wilen}},\
  }\href@noop {} {\emph {\bibinfo {title} {Stereochemistry of Organic
  Compounds}}}\ (\bibinfo  {publisher} {Wiley},\ \bibinfo {year}
  {2008})\BibitemShut {NoStop}%
\bibitem [{\citenamefont {Kelly}, \citenamefont {Jess},\ and\ \citenamefont
  {Price}(2005)}]{KELLY2005}%
  \BibitemOpen
  \bibfield  {author} {\bibinfo {author} {\bibfnamefont {S.~M.}\ \bibnamefont
  {Kelly}}, \bibinfo {author} {\bibfnamefont {T.~J.}\ \bibnamefont {Jess}}, \
  and\ \bibinfo {author} {\bibfnamefont {N.~C.}\ \bibnamefont {Price}},\ }\href
  {\doibase https://doi.org/10.1016/j.bbapap.2005.06.005} {\bibfield  {journal}
  {\bibinfo  {journal} {Biochim. Biophys. Acta}\ }\textbf {\bibinfo {volume}
  {1751}},\ \bibinfo {pages} {119 } (\bibinfo {year} {2005})}\BibitemShut
  {NoStop}%
\bibitem [{\citenamefont {Fortson}\ and\ \citenamefont
  {Lewis}(1984)}]{FORTSON1984}%
  \BibitemOpen
  \bibfield  {author} {\bibinfo {author} {\bibfnamefont {E.}~\bibnamefont
  {Fortson}}\ and\ \bibinfo {author} {\bibfnamefont {L.}~\bibnamefont
  {Lewis}},\ }\href {\doibase https://doi.org/10.1016/0370-1573(84)90005-X}
  {\bibfield  {journal} {\bibinfo  {journal} {Physics Reports}\ }\textbf
  {\bibinfo {volume} {113}},\ \bibinfo {pages} {289 } (\bibinfo {year}
  {1984})}\BibitemShut {NoStop}%
\bibitem [{\citenamefont {Budker}\ and\ \citenamefont
  {Jackson~K.}(2013)}]{budkerBook}%
  \BibitemOpen
  \bibfield  {author} {\bibinfo {author} {\bibfnamefont {D.}~\bibnamefont
  {Budker}}\ and\ \bibinfo {author} {\bibfnamefont {D.}~\bibnamefont
  {Jackson~K.}},\ }\href {\doibase 10.1017/CBO9780511846380} {\emph {\bibinfo
  {title} {Optical Magnetometry}}}\ (\bibinfo  {publisher} {Cambridge
  University Press},\ \bibinfo {year} {2013})\BibitemShut {NoStop}%
\bibitem [{\citenamefont {Busch}\ and\ \citenamefont
  {Busch}(2006)}]{Busch2006}%
  \BibitemOpen
  \bibfield  {author} {\bibinfo {author} {\bibfnamefont {K.~W.}\ \bibnamefont
  {Busch}}\ and\ \bibinfo {author} {\bibfnamefont {M.~A.}\ \bibnamefont
  {Busch}},\ }\href {\doibase 10.1016/b978-0-444-51669-5.x5000-5} {\emph
  {\bibinfo {title} {Chiral Analysis}}}\ (\bibinfo  {publisher} {Elsevier},\
  \bibinfo {year} {2006})\BibitemShut {NoStop}%
\bibitem [{\citenamefont {Argentine}, \citenamefont {Owens},\ and\
  \citenamefont {Olsen}(2007)}]{ARGENTINE2007}%
  \BibitemOpen
  \bibfield  {author} {\bibinfo {author} {\bibfnamefont {M.~D.}\ \bibnamefont
  {Argentine}}, \bibinfo {author} {\bibfnamefont {P.~K.}\ \bibnamefont
  {Owens}}, \ and\ \bibinfo {author} {\bibfnamefont {B.~A.}\ \bibnamefont
  {Olsen}},\ }\href {\doibase https://doi.org/10.1016/j.addr.2006.10.005}
  {\bibfield  {journal} {\bibinfo  {journal} {Advanced Drug Delivery Reviews}\
  }\textbf {\bibinfo {volume} {59}},\ \bibinfo {pages} {12 } (\bibinfo {year}
  {2007})},\ \bibinfo {note} {pharmaceutical Impurities: Analytical,
  Toxicological and Regulatory Perspectives}\BibitemShut {NoStop}%
\bibitem [{\citenamefont {Budker}, \citenamefont {Kimball},\ and\ \citenamefont
  {DeMille}(2008)}]{Budker2008}%
  \BibitemOpen
  \bibfield  {author} {\bibinfo {author} {\bibfnamefont {D.}~\bibnamefont
  {Budker}}, \bibinfo {author} {\bibfnamefont {D.~F.}\ \bibnamefont {Kimball}},
  \ and\ \bibinfo {author} {\bibfnamefont {D.~P.}\ \bibnamefont {DeMille}},\
  }\href {https://search.library.wisc.edu/catalog/9910057254002121} {\emph
  {\bibinfo {title} {Atomic physics : an exploration through problems and
  solutions}}},\ \bibinfo {edition} {2nd}\ ed.\ (\bibinfo  {publisher} {Oxford
  University Press},\ \bibinfo {year} {2008})\BibitemShut {NoStop}%
\bibitem [{\citenamefont {Li}\ \emph {et~al.}(2011)\citenamefont {Li},
  \citenamefont {Vachaspati}, \citenamefont {Sheng}, \citenamefont {Dural},\
  and\ \citenamefont {Romalis}}]{Li2011}%
  \BibitemOpen
  \bibfield  {author} {\bibinfo {author} {\bibfnamefont {S.}~\bibnamefont
  {Li}}, \bibinfo {author} {\bibfnamefont {P.}~\bibnamefont {Vachaspati}},
  \bibinfo {author} {\bibfnamefont {D.}~\bibnamefont {Sheng}}, \bibinfo
  {author} {\bibfnamefont {N.}~\bibnamefont {Dural}}, \ and\ \bibinfo {author}
  {\bibfnamefont {M.~V.}\ \bibnamefont {Romalis}},\ }\href {\doibase
  10.1103/PhysRevA.84.061403} {\bibfield  {journal} {\bibinfo  {journal} {Phys.
  Rev. A}\ }\textbf {\bibinfo {volume} {84}},\ \bibinfo {pages} {061403}
  (\bibinfo {year} {2011})}\BibitemShut {NoStop}%
\bibitem [{\citenamefont {Tuzson}\ \emph {et~al.}(2013)\citenamefont {Tuzson},
  \citenamefont {Mangold}, \citenamefont {Looser}, \citenamefont {Manninen},\
  and\ \citenamefont {Emmenegger}}]{Tuzson2013}%
  \BibitemOpen
  \bibfield  {author} {\bibinfo {author} {\bibfnamefont {B.}~\bibnamefont
  {Tuzson}}, \bibinfo {author} {\bibfnamefont {M.}~\bibnamefont {Mangold}},
  \bibinfo {author} {\bibfnamefont {H.}~\bibnamefont {Looser}}, \bibinfo
  {author} {\bibfnamefont {A.}~\bibnamefont {Manninen}}, \ and\ \bibinfo
  {author} {\bibfnamefont {L.}~\bibnamefont {Emmenegger}},\ }\href {\doibase
  10.1364/OL.38.000257} {\bibfield  {journal} {\bibinfo  {journal} {Opt.
  Lett.}\ }\textbf {\bibinfo {volume} {38}},\ \bibinfo {pages} {257} (\bibinfo
  {year} {2013})}\BibitemShut {NoStop}%
\bibitem [{\citenamefont {White}(1942)}]{White1942}%
  \BibitemOpen
  \bibfield  {author} {\bibinfo {author} {\bibfnamefont {J.~U.}\ \bibnamefont
  {White}},\ }\href {\doibase 10.1364/JOSA.32.000285} {\bibfield  {journal}
  {\bibinfo  {journal} {J. Opt. Soc. Am.}\ }\textbf {\bibinfo {volume} {32}},\
  \bibinfo {pages} {285} (\bibinfo {year} {1942})}\BibitemShut {NoStop}%
\bibitem [{\citenamefont {Herriott}, \citenamefont {Kogelnik},\ and\
  \citenamefont {Kompfner}(1964)}]{Herriott1964}%
  \BibitemOpen
  \bibfield  {author} {\bibinfo {author} {\bibfnamefont {D.}~\bibnamefont
  {Herriott}}, \bibinfo {author} {\bibfnamefont {H.}~\bibnamefont {Kogelnik}},
  \ and\ \bibinfo {author} {\bibfnamefont {R.}~\bibnamefont {Kompfner}},\
  }\href {\doibase 10.1364/AO.3.000523} {\bibfield  {journal} {\bibinfo
  {journal} {Appl. Opt.}\ }\textbf {\bibinfo {volume} {3}},\ \bibinfo {pages}
  {523} (\bibinfo {year} {1964})}\BibitemShut {NoStop}%
\bibitem [{\citenamefont {Das}\ and\ \citenamefont {Wilson}(2011)}]{Das2011}%
  \BibitemOpen
  \bibfield  {author} {\bibinfo {author} {\bibfnamefont {D.}~\bibnamefont
  {Das}}\ and\ \bibinfo {author} {\bibfnamefont {A.~C.}\ \bibnamefont
  {Wilson}},\ }\href {\doibase 10.1007/s00340-010-4337-7} {\bibfield  {journal}
  {\bibinfo  {journal} {Applied Physics B}\ }\textbf {\bibinfo {volume}
  {103}},\ \bibinfo {pages} {749} (\bibinfo {year} {2011})}\BibitemShut
  {NoStop}%
\bibitem [{\citenamefont {Krzempek}\ \emph {et~al.}(2013)\citenamefont
  {Krzempek}, \citenamefont {Jahjah}, \citenamefont {Lewicki}, \citenamefont
  {Stefa{\'{n}}ski}, \citenamefont {So}, \citenamefont {Thomazy},\ and\
  \citenamefont {Tittel}}]{Krzempek2013}%
  \BibitemOpen
  \bibfield  {author} {\bibinfo {author} {\bibfnamefont {K.}~\bibnamefont
  {Krzempek}}, \bibinfo {author} {\bibfnamefont {M.}~\bibnamefont {Jahjah}},
  \bibinfo {author} {\bibfnamefont {R.}~\bibnamefont {Lewicki}}, \bibinfo
  {author} {\bibfnamefont {P.}~\bibnamefont {Stefa{\'{n}}ski}}, \bibinfo
  {author} {\bibfnamefont {S.}~\bibnamefont {So}}, \bibinfo {author}
  {\bibfnamefont {D.}~\bibnamefont {Thomazy}}, \ and\ \bibinfo {author}
  {\bibfnamefont {F.~K.}\ \bibnamefont {Tittel}},\ }\href {\doibase
  10.1007/s00340-013-5544-9} {\bibfield  {journal} {\bibinfo  {journal}
  {Applied Physics B}\ }\textbf {\bibinfo {volume} {112}},\ \bibinfo {pages}
  {461} (\bibinfo {year} {2013})}\BibitemShut {NoStop}%
\bibitem [{\citenamefont {Barron}(2004)}]{barron_2004}%
  \BibitemOpen
  \bibfield  {author} {\bibinfo {author} {\bibfnamefont {L.~D.}\ \bibnamefont
  {Barron}},\ }\href {\doibase 10.1017/CBO9780511535468} {\emph {\bibinfo
  {title} {Molecular Light Scattering and Optical Activity}}},\ \bibinfo
  {edition} {2nd}\ ed.\ (\bibinfo  {publisher} {Cambridge University Press},\
  \bibinfo {year} {2004})\BibitemShut {NoStop}%
\bibitem [{\citenamefont {M\"uller}, \citenamefont {Wiberg},\ and\
  \citenamefont {Vaccaro}(2000)}]{Mueller2000}%
  \BibitemOpen
  \bibfield  {author} {\bibinfo {author} {\bibfnamefont {T.}~\bibnamefont
  {M\"uller}}, \bibinfo {author} {\bibfnamefont {K.~B.}\ \bibnamefont
  {Wiberg}}, \ and\ \bibinfo {author} {\bibfnamefont {P.~H.}\ \bibnamefont
  {Vaccaro}},\ }\href {\doibase 10.1021/jp000705n} {\bibfield  {journal}
  {\bibinfo  {journal} {The Journal of Physical Chemistry A}\ }\textbf
  {\bibinfo {volume} {104}},\ \bibinfo {pages} {5959} (\bibinfo {year}
  {2000})}\BibitemShut {NoStop}%
\bibitem [{\citenamefont {Bougas}\ \emph {et~al.}(2012)\citenamefont {Bougas},
  \citenamefont {Katsoprinakis}, \citenamefont {von Klitzing}, \citenamefont
  {Sapirstein},\ and\ \citenamefont {Rakitzis}}]{bougas2012}%
  \BibitemOpen
  \bibfield  {author} {\bibinfo {author} {\bibfnamefont {L.}~\bibnamefont
  {Bougas}}, \bibinfo {author} {\bibfnamefont {G.~E.}\ \bibnamefont
  {Katsoprinakis}}, \bibinfo {author} {\bibfnamefont {W.}~\bibnamefont {von
  Klitzing}}, \bibinfo {author} {\bibfnamefont {J.}~\bibnamefont {Sapirstein}},
  \ and\ \bibinfo {author} {\bibfnamefont {T.~P.}\ \bibnamefont {Rakitzis}},\
  }\href {\doibase 10.1103/PhysRevLett.108.210801} {\bibfield  {journal}
  {\bibinfo  {journal} {Phys. Rev. Lett.}\ }\textbf {\bibinfo {volume} {108}},\
  \bibinfo {pages} {210801} (\bibinfo {year} {2012})}\BibitemShut {NoStop}%
\bibitem [{\citenamefont {Durand}, \citenamefont {Morville},\ and\
  \citenamefont {Romanini}(2010)}]{Durand2010}%
  \BibitemOpen
  \bibfield  {author} {\bibinfo {author} {\bibfnamefont {M.}~\bibnamefont
  {Durand}}, \bibinfo {author} {\bibfnamefont {J.}~\bibnamefont {Morville}}, \
  and\ \bibinfo {author} {\bibfnamefont {D.}~\bibnamefont {Romanini}},\ }\href
  {\doibase 10.1103/PhysRevA.82.031803} {\bibfield  {journal} {\bibinfo
  {journal} {Phys. Rev. A}\ }\textbf {\bibinfo {volume} {82}},\ \bibinfo
  {pages} {031803} (\bibinfo {year} {2010})}\BibitemShut {NoStop}%
\bibitem [{\citenamefont {Gianella}\ \emph {et~al.}(2017)\citenamefont
  {Gianella}, \citenamefont {Pinto}, \citenamefont {Wu},\ and\ \citenamefont
  {Ritchie}}]{Gianella2017}%
  \BibitemOpen
  \bibfield  {author} {\bibinfo {author} {\bibfnamefont {M.}~\bibnamefont
  {Gianella}}, \bibinfo {author} {\bibfnamefont {T.~H.~P.}\ \bibnamefont
  {Pinto}}, \bibinfo {author} {\bibfnamefont {X.}~\bibnamefont {Wu}}, \ and\
  \bibinfo {author} {\bibfnamefont {G.~A.~D.}\ \bibnamefont {Ritchie}},\ }\href
  {\doibase 10.1063/1.4985900} {\bibfield  {journal} {\bibinfo  {journal} {The
  Journal of Chemical Physics}\ }\textbf {\bibinfo {volume} {147}},\ \bibinfo
  {pages} {054201} (\bibinfo {year} {2017})}\BibitemShut {NoStop}%
\bibitem [{\citenamefont {Hall}, \citenamefont {Ye},\ and\ \citenamefont
  {Ma}(2000)}]{Hall2000}%
  \BibitemOpen
  \bibfield  {author} {\bibinfo {author} {\bibfnamefont {J.~L.}\ \bibnamefont
  {Hall}}, \bibinfo {author} {\bibfnamefont {J.}~\bibnamefont {Ye}}, \ and\
  \bibinfo {author} {\bibfnamefont {L.-S.}\ \bibnamefont {Ma}},\ }\href
  {\doibase 10.1103/PhysRevA.62.013815} {\bibfield  {journal} {\bibinfo
  {journal} {Phys. Rev. A}\ }\textbf {\bibinfo {volume} {62}},\ \bibinfo
  {pages} {013815} (\bibinfo {year} {2000})}\BibitemShut {NoStop}%
\bibitem [{\citenamefont {Bailly}, \citenamefont {Thon},\ and\ \citenamefont
  {Robilliard}(2010)}]{Gilles2010}%
  \BibitemOpen
  \bibfield  {author} {\bibinfo {author} {\bibfnamefont {G.}~\bibnamefont
  {Bailly}}, \bibinfo {author} {\bibfnamefont {R.}~\bibnamefont {Thon}}, \ and\
  \bibinfo {author} {\bibfnamefont {C.}~\bibnamefont {Robilliard}},\ }\href
  {\doibase 10.1063/1.3356731} {\bibfield  {journal} {\bibinfo  {journal}
  {Review of Scientific Instruments}\ }\textbf {\bibinfo {volume} {81}},\
  \bibinfo {pages} {033105} (\bibinfo {year} {2010})}\BibitemShut {NoStop}%
\bibitem [{\citenamefont {Wheeler}\ \emph {et~al.}(1998)\citenamefont
  {Wheeler}, \citenamefont {Newman}, \citenamefont {Orr-Ewing},\ and\
  \citenamefont {Ashfold}}]{Wheeler1998}%
  \BibitemOpen
  \bibfield  {author} {\bibinfo {author} {\bibfnamefont {M.~D.}\ \bibnamefont
  {Wheeler}}, \bibinfo {author} {\bibfnamefont {S.~M.}\ \bibnamefont {Newman}},
  \bibinfo {author} {\bibfnamefont {A.~J.}\ \bibnamefont {Orr-Ewing}}, \ and\
  \bibinfo {author} {\bibfnamefont {M.~N.~R.}\ \bibnamefont {Ashfold}},\ }\href
  {\doibase 10.1039/A707686J} {\bibfield  {journal} {\bibinfo  {journal} {J.
  Chem. Soc.{,} Faraday Trans.}\ }\textbf {\bibinfo {volume} {94}},\ \bibinfo
  {pages} {337} (\bibinfo {year} {1998})}\BibitemShut {NoStop}%
\bibitem [{\citenamefont {Berden}\ and\ \citenamefont
  {Engeln}(2009)}]{Berden2009}%
  \BibitemOpen
  \bibfield  {author} {\bibinfo {author} {\bibfnamefont {G.}~\bibnamefont
  {Berden}}\ and\ \bibinfo {author} {\bibfnamefont {R.}~\bibnamefont
  {Engeln}},\ }\href@noop {} {\emph {\bibinfo {title} {Cavity Ring-Down
  Spectroscopy: Techniques and Applications}}}\ (\bibinfo  {publisher}
  {Wiley},\ \bibinfo {year} {2009})\BibitemShut {NoStop}%
\bibitem [{\citenamefont {Engeln}\ \emph {et~al.}(1997)\citenamefont {Engeln},
  \citenamefont {Berden}, \citenamefont {van~den Berg},\ and\ \citenamefont
  {Meijer}}]{Engeln1997}%
  \BibitemOpen
  \bibfield  {author} {\bibinfo {author} {\bibfnamefont {R.}~\bibnamefont
  {Engeln}}, \bibinfo {author} {\bibfnamefont {G.}~\bibnamefont {Berden}},
  \bibinfo {author} {\bibfnamefont {E.}~\bibnamefont {van~den Berg}}, \ and\
  \bibinfo {author} {\bibfnamefont {G.}~\bibnamefont {Meijer}},\ }\href
  {\doibase 10.1063/1.474808} {\bibfield  {journal} {\bibinfo  {journal} {The
  Journal of Chemical Physics}\ }\textbf {\bibinfo {volume} {107}},\ \bibinfo
  {pages} {4458} (\bibinfo {year} {1997})}\BibitemShut {NoStop}%
\bibitem [{\citenamefont {Engeln}\ \emph {et~al.}(1998)\citenamefont {Engeln},
  \citenamefont {Berden}, \citenamefont {Peeters},\ and\ \citenamefont
  {Meijer}}]{Engeln1998}%
  \BibitemOpen
  \bibfield  {author} {\bibinfo {author} {\bibfnamefont {R.}~\bibnamefont
  {Engeln}}, \bibinfo {author} {\bibfnamefont {G.}~\bibnamefont {Berden}},
  \bibinfo {author} {\bibfnamefont {R.}~\bibnamefont {Peeters}}, \ and\
  \bibinfo {author} {\bibfnamefont {G.}~\bibnamefont {Meijer}},\ }\href
  {\doibase 10.1063/1.1149176} {\bibfield  {journal} {\bibinfo  {journal}
  {Review of Scientific Instruments}\ }\textbf {\bibinfo {volume} {69}},\
  \bibinfo {pages} {3763} (\bibinfo {year} {1998})},\ \Eprint
  {http://arxiv.org/abs/https://doi.org/10.1063/1.1149176}
  {https://doi.org/10.1063/1.1149176} \BibitemShut {NoStop}%
\bibitem [{\citenamefont {Lee}\ \emph {et~al.}(2000)\citenamefont {Lee},
  \citenamefont {Lee}, \citenamefont {Kim}, \citenamefont {Yoo},\ and\
  \citenamefont {Hahn}}]{YongLee2000}%
  \BibitemOpen
  \bibfield  {author} {\bibinfo {author} {\bibfnamefont {J.~Y.}\ \bibnamefont
  {Lee}}, \bibinfo {author} {\bibfnamefont {H.-W.}\ \bibnamefont {Lee}},
  \bibinfo {author} {\bibfnamefont {J.~W.}\ \bibnamefont {Kim}}, \bibinfo
  {author} {\bibfnamefont {Y.~S.}\ \bibnamefont {Yoo}}, \ and\ \bibinfo
  {author} {\bibfnamefont {J.~W.}\ \bibnamefont {Hahn}},\ }\href {\doibase
  10.1364/AO.39.001941} {\bibfield  {journal} {\bibinfo  {journal} {Appl.
  Opt.}\ }\textbf {\bibinfo {volume} {39}},\ \bibinfo {pages} {1941} (\bibinfo
  {year} {2000})}\BibitemShut {NoStop}%
\bibitem [{\citenamefont {Huang}\ and\ \citenamefont
  {Lehmann}(2008)}]{Huang2008}%
  \BibitemOpen
  \bibfield  {author} {\bibinfo {author} {\bibfnamefont {H.}~\bibnamefont
  {Huang}}\ and\ \bibinfo {author} {\bibfnamefont {K.~K.}\ \bibnamefont
  {Lehmann}},\ }\href {\doibase 10.1364/AO.47.003817} {\bibfield  {journal}
  {\bibinfo  {journal} {Appl. Opt.}\ }\textbf {\bibinfo {volume} {47}},\
  \bibinfo {pages} {3817} (\bibinfo {year} {2008})}\BibitemShut {NoStop}%
\bibitem [{\citenamefont {Dupr\'e}(2015)}]{Dupre2015}%
  \BibitemOpen
  \bibfield  {author} {\bibinfo {author} {\bibfnamefont {P.}~\bibnamefont
  {Dupr\'e}},\ }\href {\doibase 10.1103/PhysRevA.92.053817} {\bibfield
  {journal} {\bibinfo  {journal} {Phys. Rev. A}\ }\textbf {\bibinfo {volume}
  {92}},\ \bibinfo {pages} {053817} (\bibinfo {year} {2015})}\BibitemShut
  {NoStop}%
\bibitem [{\citenamefont {Fleisher}\ \emph {et~al.}(2016)\citenamefont
  {Fleisher}, \citenamefont {Long}, \citenamefont {Liu},\ and\ \citenamefont
  {Hodges}}]{Fleisher2016}%
  \BibitemOpen
  \bibfield  {author} {\bibinfo {author} {\bibfnamefont {A.~J.}\ \bibnamefont
  {Fleisher}}, \bibinfo {author} {\bibfnamefont {D.~A.}\ \bibnamefont {Long}},
  \bibinfo {author} {\bibfnamefont {Q.}~\bibnamefont {Liu}}, \ and\ \bibinfo
  {author} {\bibfnamefont {J.~T.}\ \bibnamefont {Hodges}},\ }\href {\doibase
  10.1103/PhysRevA.93.013833} {\bibfield  {journal} {\bibinfo  {journal} {Phys.
  Rev. A}\ }\textbf {\bibinfo {volume} {93}},\ \bibinfo {pages} {013833}
  (\bibinfo {year} {2016})}\BibitemShut {NoStop}%
\bibitem [{\citenamefont {Berden}\ \emph {et~al.}(1998)\citenamefont {Berden},
  \citenamefont {Engeln}, \citenamefont {Christianen}, \citenamefont {Maan},\
  and\ \citenamefont {Meijer}}]{Berden1998}%
  \BibitemOpen
  \bibfield  {author} {\bibinfo {author} {\bibfnamefont {G.}~\bibnamefont
  {Berden}}, \bibinfo {author} {\bibfnamefont {R.}~\bibnamefont {Engeln}},
  \bibinfo {author} {\bibfnamefont {P.~C.~M.}\ \bibnamefont {Christianen}},
  \bibinfo {author} {\bibfnamefont {J.~C.}\ \bibnamefont {Maan}}, \ and\
  \bibinfo {author} {\bibfnamefont {G.}~\bibnamefont {Meijer}},\ }\href
  {\doibase 10.1103/PhysRevA.58.3114} {\bibfield  {journal} {\bibinfo
  {journal} {Phys. Rev. A}\ }\textbf {\bibinfo {volume} {58}},\ \bibinfo
  {pages} {3114} (\bibinfo {year} {1998})}\BibitemShut {NoStop}%
\bibitem [{\citenamefont {Hayden}\ \emph {et~al.}(2018)\citenamefont {Hayden},
  \citenamefont {Westberg}, \citenamefont {Patrick}, \citenamefont {Lendl},\
  and\ \citenamefont {Wysocki}}]{Hayden2018}%
  \BibitemOpen
  \bibfield  {author} {\bibinfo {author} {\bibfnamefont {J.}~\bibnamefont
  {Hayden}}, \bibinfo {author} {\bibfnamefont {J.}~\bibnamefont {Westberg}},
  \bibinfo {author} {\bibfnamefont {C.~L.}\ \bibnamefont {Patrick}}, \bibinfo
  {author} {\bibfnamefont {B.}~\bibnamefont {Lendl}}, \ and\ \bibinfo {author}
  {\bibfnamefont {G.}~\bibnamefont {Wysocki}},\ }\href {\doibase
  10.1364/OL.43.005046} {\bibfield  {journal} {\bibinfo  {journal} {Opt.
  Lett.}\ }\textbf {\bibinfo {volume} {43}},\ \bibinfo {pages} {5046} (\bibinfo
  {year} {2018})}\BibitemShut {NoStop}%
\bibitem [{\citenamefont {Patrick}, \citenamefont {Westberg},\ and\
  \citenamefont {Wysocki}(2019)}]{Patrick2019}%
  \BibitemOpen
  \bibfield  {author} {\bibinfo {author} {\bibfnamefont {C.~L.}\ \bibnamefont
  {Patrick}}, \bibinfo {author} {\bibfnamefont {J.}~\bibnamefont {Westberg}}, \
  and\ \bibinfo {author} {\bibfnamefont {G.}~\bibnamefont {Wysocki}},\ }\href
  {\doibase 10.1021/acs.analchem.8b04359} {\bibfield  {journal} {\bibinfo
  {journal} {Analytical Chemistry}\ }\textbf {\bibinfo {volume} {91}},\
  \bibinfo {pages} {1696} (\bibinfo {year} {2019})}\BibitemShut {NoStop}%
\bibitem [{\citenamefont {Gianella}\ \emph {et~al.}(2019)\citenamefont
  {Gianella}, \citenamefont {Press}, \citenamefont {Manfred}, \citenamefont
  {Norman}, \citenamefont {Islam},\ and\ \citenamefont
  {Ritchie}}]{Gianella2019}%
  \BibitemOpen
  \bibfield  {author} {\bibinfo {author} {\bibfnamefont {M.}~\bibnamefont
  {Gianella}}, \bibinfo {author} {\bibfnamefont {S.~A.}\ \bibnamefont {Press}},
  \bibinfo {author} {\bibfnamefont {K.~M.}\ \bibnamefont {Manfred}}, \bibinfo
  {author} {\bibfnamefont {H.~C.}\ \bibnamefont {Norman}}, \bibinfo {author}
  {\bibfnamefont {M.}~\bibnamefont {Islam}}, \ and\ \bibinfo {author}
  {\bibfnamefont {G.~A.~D.}\ \bibnamefont {Ritchie}},\ }\href {\doibase
  10.1063/1.5119191} {\bibfield  {journal} {\bibinfo  {journal} {The Journal of
  Chemical Physics}\ }\textbf {\bibinfo {volume} {151}},\ \bibinfo {pages}
  {124202} (\bibinfo {year} {2019})}\BibitemShut {NoStop}%
\bibitem [{\citenamefont {Gianella}\ \emph {et~al.}(2018)\citenamefont
  {Gianella}, \citenamefont {Reuter}, \citenamefont {Press}, \citenamefont
  {Schmidt-Bleker}, \citenamefont {van Helden},\ and\ \citenamefont
  {Ritchie}}]{Gianella2018}%
  \BibitemOpen
  \bibfield  {author} {\bibinfo {author} {\bibfnamefont {M.}~\bibnamefont
  {Gianella}}, \bibinfo {author} {\bibfnamefont {S.}~\bibnamefont {Reuter}},
  \bibinfo {author} {\bibfnamefont {S.~A.}\ \bibnamefont {Press}}, \bibinfo
  {author} {\bibfnamefont {A.}~\bibnamefont {Schmidt-Bleker}}, \bibinfo
  {author} {\bibfnamefont {J.~H.}\ \bibnamefont {van Helden}}, \ and\ \bibinfo
  {author} {\bibfnamefont {G.~A.~D.}\ \bibnamefont {Ritchie}},\ }\href
  {\doibase 10.1088/1361-6595/aadf01} {\bibfield  {journal} {\bibinfo
  {journal} {Plasma Sources Science and Technology}\ }\textbf {\bibinfo
  {volume} {27}},\ \bibinfo {pages} {095013} (\bibinfo {year}
  {2018})}\BibitemShut {NoStop}%
\bibitem [{\citenamefont {M\"{u}ller}\ \emph {et~al.}(2002)\citenamefont
  {M\"{u}ller}, \citenamefont {Wiberg}, \citenamefont {Vaccaro}, \citenamefont
  {Cheeseman},\ and\ \citenamefont {Frisch}}]{Muller2002}%
  \BibitemOpen
  \bibfield  {author} {\bibinfo {author} {\bibfnamefont {T.}~\bibnamefont
  {M\"{u}ller}}, \bibinfo {author} {\bibfnamefont {K.~B.}\ \bibnamefont
  {Wiberg}}, \bibinfo {author} {\bibfnamefont {P.~H.}\ \bibnamefont {Vaccaro}},
  \bibinfo {author} {\bibfnamefont {J.~R.}\ \bibnamefont {Cheeseman}}, \ and\
  \bibinfo {author} {\bibfnamefont {M.~J.}\ \bibnamefont {Frisch}},\ }\href
  {\doibase 10.1364/JOSAB.19.000125} {\bibfield  {journal} {\bibinfo  {journal}
  {J. Opt. Soc. Am. B}\ }\textbf {\bibinfo {volume} {19}},\ \bibinfo {pages}
  {125} (\bibinfo {year} {2002})}\BibitemShut {NoStop}%
\bibitem [{\citenamefont {Sofikitis}\ \emph {et~al.}(2014)\citenamefont
  {Sofikitis}, \citenamefont {Bougas}, \citenamefont {Katsoprinakis},
  \citenamefont {Spiliotis}, \citenamefont {Loppinet},\ and\ \citenamefont
  {Rakitzis}}]{Sofikitis2014}%
  \BibitemOpen
  \bibfield  {author} {\bibinfo {author} {\bibfnamefont {D.}~\bibnamefont
  {Sofikitis}}, \bibinfo {author} {\bibfnamefont {L.}~\bibnamefont {Bougas}},
  \bibinfo {author} {\bibfnamefont {G.~E.}\ \bibnamefont {Katsoprinakis}},
  \bibinfo {author} {\bibfnamefont {A.~K.}\ \bibnamefont {Spiliotis}}, \bibinfo
  {author} {\bibfnamefont {B.}~\bibnamefont {Loppinet}}, \ and\ \bibinfo
  {author} {\bibfnamefont {T.~P.}\ \bibnamefont {Rakitzis}},\ }\href
  {http://dx.doi.org/10.1038/nature13680 http://10.0.4.14/nature13680}
  {\bibfield  {journal} {\bibinfo  {journal} {Nature}\ }\textbf {\bibinfo
  {volume} {514}},\ \bibinfo {pages} {76} (\bibinfo {year} {2014})}\BibitemShut
  {NoStop}%
\bibitem [{\citenamefont {Bougas}\ \emph {et~al.}(2015)\citenamefont {Bougas},
  \citenamefont {Sofikitis}, \citenamefont {Katsoprinakis}, \citenamefont
  {Spiliotis}, \citenamefont {Tzallas}, \citenamefont {Loppinet},\ and\
  \citenamefont {Rakitzis}}]{Bougas2015}%
  \BibitemOpen
  \bibfield  {author} {\bibinfo {author} {\bibfnamefont {L.}~\bibnamefont
  {Bougas}}, \bibinfo {author} {\bibfnamefont {D.}~\bibnamefont {Sofikitis}},
  \bibinfo {author} {\bibfnamefont {G.~E.}\ \bibnamefont {Katsoprinakis}},
  \bibinfo {author} {\bibfnamefont {A.~K.}\ \bibnamefont {Spiliotis}}, \bibinfo
  {author} {\bibfnamefont {P.}~\bibnamefont {Tzallas}}, \bibinfo {author}
  {\bibfnamefont {B.}~\bibnamefont {Loppinet}}, \ and\ \bibinfo {author}
  {\bibfnamefont {T.~P.}\ \bibnamefont {Rakitzis}},\ }\href {\doibase
  10.1063/1.4930109} {\bibfield  {journal} {\bibinfo  {journal} {The Journal of
  Chemical Physics}\ }\textbf {\bibinfo {volume} {143}},\ \bibinfo {pages}
  {104202} (\bibinfo {year} {2015})}\BibitemShut {NoStop}%
\bibitem [{\citenamefont {Romanini}\ \emph {et~al.}(1997)\citenamefont
  {Romanini}, \citenamefont {Kachanov}, \citenamefont {Sadeghi},\ and\
  \citenamefont {Stoeckel}}]{ROMANINI1997}%
  \BibitemOpen
  \bibfield  {author} {\bibinfo {author} {\bibfnamefont {D.}~\bibnamefont
  {Romanini}}, \bibinfo {author} {\bibfnamefont {A.}~\bibnamefont {Kachanov}},
  \bibinfo {author} {\bibfnamefont {N.}~\bibnamefont {Sadeghi}}, \ and\
  \bibinfo {author} {\bibfnamefont {F.}~\bibnamefont {Stoeckel}},\ }\href
  {\doibase https://doi.org/10.1016/S0009-2614(96)01351-6} {\bibfield
  {journal} {\bibinfo  {journal} {Chemical Physics Letters}\ }\textbf {\bibinfo
  {volume} {264}},\ \bibinfo {pages} {316 } (\bibinfo {year}
  {1997})}\BibitemShut {NoStop}%
\bibitem [{Note1()}]{Note1}%
  \BibitemOpen
  \bibinfo {note} {Los Gatos Research, http://www.lgrinc.com; Picarro,
  https://www.picarro.com}\BibitemShut {NoStop}%
\bibitem [{\citenamefont {{Nilsson}}, \citenamefont {{Gustafson}},\ and\
  \citenamefont {{Byer}}(1989)}]{Nilsson1989}%
  \BibitemOpen
  \bibfield  {author} {\bibinfo {author} {\bibfnamefont {A.~C.}\ \bibnamefont
  {{Nilsson}}}, \bibinfo {author} {\bibfnamefont {E.~K.}\ \bibnamefont
  {{Gustafson}}}, \ and\ \bibinfo {author} {\bibfnamefont {R.~L.}\ \bibnamefont
  {{Byer}}},\ }\href {\doibase 10.1109/3.17343} {\bibfield  {journal} {\bibinfo
   {journal} {IEEE Journal of Quantum Electronics}\ }\textbf {\bibinfo {volume}
  {25}},\ \bibinfo {pages} {767} (\bibinfo {year} {1989})}\BibitemShut
  {NoStop}%
\bibitem [{\citenamefont {Vallet}\ \emph {et~al.}(1999)\citenamefont {Vallet},
  \citenamefont {Bretenaker}, \citenamefont {Floch}, \citenamefont {Naour},\
  and\ \citenamefont {Oger}}]{VALLET1999}%
  \BibitemOpen
  \bibfield  {author} {\bibinfo {author} {\bibfnamefont {M.}~\bibnamefont
  {Vallet}}, \bibinfo {author} {\bibfnamefont {F.}~\bibnamefont {Bretenaker}},
  \bibinfo {author} {\bibfnamefont {A.~L.}\ \bibnamefont {Floch}}, \bibinfo
  {author} {\bibfnamefont {R.~L.}\ \bibnamefont {Naour}}, \ and\ \bibinfo
  {author} {\bibfnamefont {M.}~\bibnamefont {Oger}},\ }\href {\doibase
  https://doi.org/10.1016/S0030-4018(99)00351-X} {\bibfield  {journal}
  {\bibinfo  {journal} {Optics Communications}\ }\textbf {\bibinfo {volume}
  {168}},\ \bibinfo {pages} {423 } (\bibinfo {year} {1999})}\BibitemShut
  {NoStop}%
\bibitem [{\citenamefont {Bougas}\ \emph {et~al.}(2014)\citenamefont {Bougas},
  \citenamefont {Katsoprinakis}, \citenamefont {von Klitzing},\ and\
  \citenamefont {Rakitzis}}]{Bougas2014}%
  \BibitemOpen
  \bibfield  {author} {\bibinfo {author} {\bibfnamefont {L.}~\bibnamefont
  {Bougas}}, \bibinfo {author} {\bibfnamefont {G.~E.}\ \bibnamefont
  {Katsoprinakis}}, \bibinfo {author} {\bibfnamefont {W.}~\bibnamefont {von
  Klitzing}}, \ and\ \bibinfo {author} {\bibfnamefont {T.~P.}\ \bibnamefont
  {Rakitzis}},\ }\href {\doibase 10.1103/PhysRevA.89.052127} {\bibfield
  {journal} {\bibinfo  {journal} {Phys. Rev. A}\ }\textbf {\bibinfo {volume}
  {89}},\ \bibinfo {pages} {052127} (\bibinfo {year} {2014})}\BibitemShut
  {NoStop}%
\bibitem [{\citenamefont {Budker}\ \emph {et~al.}(2002)\citenamefont {Budker},
  \citenamefont {Gawlik}, \citenamefont {Kimball}, \citenamefont {Rochester},
  \citenamefont {Yashchuk},\ and\ \citenamefont {Weis}}]{BudkerRMP}%
  \BibitemOpen
  \bibfield  {author} {\bibinfo {author} {\bibfnamefont {D.}~\bibnamefont
  {Budker}}, \bibinfo {author} {\bibfnamefont {W.}~\bibnamefont {Gawlik}},
  \bibinfo {author} {\bibfnamefont {D.~F.}\ \bibnamefont {Kimball}}, \bibinfo
  {author} {\bibfnamefont {S.~M.}\ \bibnamefont {Rochester}}, \bibinfo {author}
  {\bibfnamefont {V.~V.}\ \bibnamefont {Yashchuk}}, \ and\ \bibinfo {author}
  {\bibfnamefont {A.}~\bibnamefont {Weis}},\ }\href {\doibase
  10.1103/RevModPhys.74.1153} {\bibfield  {journal} {\bibinfo  {journal} {Rev.
  Mod. Phys.}\ }\textbf {\bibinfo {volume} {74}},\ \bibinfo {pages} {1153}
  (\bibinfo {year} {2002})}\BibitemShut {NoStop}%
\bibitem [{\citenamefont {Sofikitis}\ \emph {et~al.}(2018)\citenamefont
  {Sofikitis}, \citenamefont {Katsoprinakis}, \citenamefont {Spiliotis},\ and\
  \citenamefont {Rakitzis}}]{SOFIKITIS2018}%
  \BibitemOpen
  \bibfield  {author} {\bibinfo {author} {\bibfnamefont {D.}~\bibnamefont
  {Sofikitis}}, \bibinfo {author} {\bibfnamefont {G.~E.}\ \bibnamefont
  {Katsoprinakis}}, \bibinfo {author} {\bibfnamefont {A.~K.}\ \bibnamefont
  {Spiliotis}}, \ and\ \bibinfo {author} {\bibfnamefont {T.~P.}\ \bibnamefont
  {Rakitzis}},\ }in\ \href {\doibase
  https://doi.org/10.1016/B978-0-444-64027-7.00018-5} {\emph {\bibinfo
  {booktitle} {Chiral Analysis (Second Edition)}}},\ \bibinfo {editor} {edited
  by\ \bibinfo {editor} {\bibfnamefont {P.~L.}\ \bibnamefont {Polavarapu}}}\
  (\bibinfo  {publisher} {Elsevier},\ \bibinfo {year} {2018})\ \bibinfo
  {edition} {second edition}\ ed.,\ pp.\ \bibinfo {pages} {649 --
  678}\BibitemShut {NoStop}%
\bibitem [{\citenamefont {Black}(2001)}]{Black2001}%
  \BibitemOpen
  \bibfield  {author} {\bibinfo {author} {\bibfnamefont {E.~D.}\ \bibnamefont
  {Black}},\ }\href {\doibase 10.1119/1.1286663} {\bibfield  {journal}
  {\bibinfo  {journal} {American Journal of Physics}\ }\textbf {\bibinfo
  {volume} {69}},\ \bibinfo {pages} {79} (\bibinfo {year} {2001})}\BibitemShut
  {NoStop}%
\bibitem [{Note2()}]{Note2}%
  \BibitemOpen
  \bibinfo {note} {Https://www.scipy.org}\BibitemShut {NoStop}%
\bibitem [{\citenamefont {Sofikitis}\ \emph {et~al.}(2015)\citenamefont
  {Sofikitis}, \citenamefont {Spiliotis}, \citenamefont {Stamataki},
  \citenamefont {Katsoprinakis}, \citenamefont {Bougas}, \citenamefont
  {Samartzis}, \citenamefont {Loppinet}, \citenamefont {Rakitzis},
  \citenamefont {Surligas},\ and\ \citenamefont
  {Papadakis}}]{Sofikitis2015SDR}%
  \BibitemOpen
  \bibfield  {author} {\bibinfo {author} {\bibfnamefont {D.}~\bibnamefont
  {Sofikitis}}, \bibinfo {author} {\bibfnamefont {A.~K.}\ \bibnamefont
  {Spiliotis}}, \bibinfo {author} {\bibfnamefont {K.}~\bibnamefont
  {Stamataki}}, \bibinfo {author} {\bibfnamefont {G.~E.}\ \bibnamefont
  {Katsoprinakis}}, \bibinfo {author} {\bibfnamefont {L.}~\bibnamefont
  {Bougas}}, \bibinfo {author} {\bibfnamefont {P.~C.}\ \bibnamefont
  {Samartzis}}, \bibinfo {author} {\bibfnamefont {B.}~\bibnamefont {Loppinet}},
  \bibinfo {author} {\bibfnamefont {T.~P.}\ \bibnamefont {Rakitzis}}, \bibinfo
  {author} {\bibfnamefont {M.}~\bibnamefont {Surligas}}, \ and\ \bibinfo
  {author} {\bibfnamefont {S.}~\bibnamefont {Papadakis}},\ }\href {\doibase
  10.1364/AO.54.005861} {\bibfield  {journal} {\bibinfo  {journal} {Appl.
  Opt.}\ }\textbf {\bibinfo {volume} {54}},\ \bibinfo {pages} {5861} (\bibinfo
  {year} {2015})}\BibitemShut {NoStop}%
\bibitem [{\citenamefont {Kitamura}, \citenamefont {Pilon},\ and\ \citenamefont
  {Jonasz}(2007)}]{Kitamura2007}%
  \BibitemOpen
  \bibfield  {author} {\bibinfo {author} {\bibfnamefont {R.}~\bibnamefont
  {Kitamura}}, \bibinfo {author} {\bibfnamefont {L.}~\bibnamefont {Pilon}}, \
  and\ \bibinfo {author} {\bibfnamefont {M.}~\bibnamefont {Jonasz}},\ }\href
  {\doibase 10.1364/AO.46.008118} {\bibfield  {journal} {\bibinfo  {journal}
  {Appl. Opt.}\ }\textbf {\bibinfo {volume} {46}},\ \bibinfo {pages} {8118}
  (\bibinfo {year} {2007})}\BibitemShut {NoStop}%
\bibitem [{\citenamefont {Ramaseshan}(1946)}]{ramaseshan1946determination}%
  \BibitemOpen
  \bibfield  {author} {\bibinfo {author} {\bibfnamefont {S.}~\bibnamefont
  {Ramaseshan}},\ }\href@noop {} {\bibfield  {journal} {\bibinfo  {journal}
  {Proceedings Mathematical Sciences}\ }\textbf {\bibinfo {volume} {24}},\
  \bibinfo {pages} {426} (\bibinfo {year} {1946})}\BibitemShut {NoStop}%
\bibitem [{\citenamefont {Ramaseshan}\ and\ \citenamefont
  {Sivaramakrishnan}(1958)}]{ramaseshan1958faraday}%
  \BibitemOpen
  \bibfield  {author} {\bibinfo {author} {\bibfnamefont {S.}~\bibnamefont
  {Ramaseshan}}\ and\ \bibinfo {author} {\bibfnamefont {V.}~\bibnamefont
  {Sivaramakrishnan}},\ }\href@noop {} {\emph {\bibinfo {title} {Faraday effect
  in diamagnetic crystals}}},\ Vol.~\bibinfo {volume} {1}\ (\bibinfo
  {publisher} {Interscience},\ \bibinfo {address} {New York},\ \bibinfo {year}
  {1958})\BibitemShut {NoStop}%
\bibitem [{\citenamefont {Shimamura}\ \emph {et~al.}(2004)\citenamefont
  {Shimamura}, \citenamefont {Vı́llora}, \citenamefont {Nakakita},
  \citenamefont {Nikl},\ and\ \citenamefont {Ichinose}}]{SHIMAMURA2004208}%
  \BibitemOpen
  \bibfield  {author} {\bibinfo {author} {\bibfnamefont {K.}~\bibnamefont
  {Shimamura}}, \bibinfo {author} {\bibfnamefont {E.~G.}\ \bibnamefont
  {Vı́llora}}, \bibinfo {author} {\bibfnamefont {S.}~\bibnamefont
  {Nakakita}}, \bibinfo {author} {\bibfnamefont {M.}~\bibnamefont {Nikl}}, \
  and\ \bibinfo {author} {\bibfnamefont {N.}~\bibnamefont {Ichinose}},\ }\href
  {\doibase https://doi.org/10.1016/j.jcrysgro.2003.12.018} {\bibfield
  {journal} {\bibinfo  {journal} {Journal of Crystal Growth}\ }\textbf
  {\bibinfo {volume} {264}},\ \bibinfo {pages} {208 } (\bibinfo {year}
  {2004})}\BibitemShut {NoStop}%
\bibitem [{\citenamefont {Molina}\ \emph {et~al.}(2011)\citenamefont {Molina},
  \citenamefont {Vasyliev}, \citenamefont {V\'{i}llora},\ and\ \citenamefont
  {Shimamura}}]{Molina2011}%
  \BibitemOpen
  \bibfield  {author} {\bibinfo {author} {\bibfnamefont {P.}~\bibnamefont
  {Molina}}, \bibinfo {author} {\bibfnamefont {V.}~\bibnamefont {Vasyliev}},
  \bibinfo {author} {\bibfnamefont {E.~G.}\ \bibnamefont {V\'{i}llora}}, \ and\
  \bibinfo {author} {\bibfnamefont {K.}~\bibnamefont {Shimamura}},\ }\href
  {\doibase 10.1364/OE.19.011786} {\bibfield  {journal} {\bibinfo  {journal}
  {Opt. Express}\ }\textbf {\bibinfo {volume} {19}},\ \bibinfo {pages} {11786}
  (\bibinfo {year} {2011})}\BibitemShut {NoStop}%
\bibitem [{\citenamefont {Ingersoll}\ and\ \citenamefont
  {Liebenberg}(1954)}]{ingersoll1954faraday}%
  \BibitemOpen
  \bibfield  {author} {\bibinfo {author} {\bibfnamefont {L.}~\bibnamefont
  {Ingersoll}}\ and\ \bibinfo {author} {\bibfnamefont {D.}~\bibnamefont
  {Liebenberg}},\ }\href@noop {} {\bibfield  {journal} {\bibinfo  {journal}
  {JOSA}\ }\textbf {\bibinfo {volume} {44}},\ \bibinfo {pages} {566} (\bibinfo
  {year} {1954})}\BibitemShut {NoStop}%
\bibitem [{\citenamefont {Ingersoll}\ and\ \citenamefont
  {Liebenberg}(1956)}]{Ingersoll1956}%
  \BibitemOpen
  \bibfield  {author} {\bibinfo {author} {\bibfnamefont {L.~R.}\ \bibnamefont
  {Ingersoll}}\ and\ \bibinfo {author} {\bibfnamefont {D.~H.}\ \bibnamefont
  {Liebenberg}},\ }\href {\doibase 10.1364/JOSA.46.000538} {\bibfield
  {journal} {\bibinfo  {journal} {J. Opt. Soc. Am.}\ }\textbf {\bibinfo
  {volume} {46}},\ \bibinfo {pages} {538} (\bibinfo {year} {1956})}\BibitemShut
  {NoStop}%
\bibitem [{\citenamefont {{Ying-Xian Yao}}\ and\ \citenamefont
  {{Pandit}}(1995)}]{Yao1995}%
  \BibitemOpen
  \bibfield  {author} {\bibinfo {author} {\bibnamefont {{Ying-Xian Yao}}}\ and\
  \bibinfo {author} {\bibfnamefont {S.~M.}\ \bibnamefont {{Pandit}}},\ }\href
  {\doibase 10.1109/78.376840} {\bibfield  {journal} {\bibinfo  {journal} {IEEE
  Transactions on Signal Processing}\ }\textbf {\bibinfo {volume} {43}},\
  \bibinfo {pages} {878} (\bibinfo {year} {1995})}\BibitemShut {NoStop}%
\bibitem [{\citenamefont {Gemmel}\ \emph {et~al.}(2010)\citenamefont {Gemmel},
  \citenamefont {Heil}, \citenamefont {Karpuk}, \citenamefont {Lenz},
  \citenamefont {Ludwig}, \citenamefont {Sobolev}, \citenamefont {Tullney},
  \citenamefont {Burghoff}, \citenamefont {Kilian}, \citenamefont
  {Knappe-Gr{\"u}neberg}, \citenamefont {M{\"u}ller}, \citenamefont {Schnabel},
  \citenamefont {Seifert}, \citenamefont {Trahms},\ and\ \citenamefont
  {Bae{\ss}ler}}]{Gemmel2010}%
  \BibitemOpen
  \bibfield  {author} {\bibinfo {author} {\bibfnamefont {C.}~\bibnamefont
  {Gemmel}}, \bibinfo {author} {\bibfnamefont {W.}~\bibnamefont {Heil}},
  \bibinfo {author} {\bibfnamefont {S.}~\bibnamefont {Karpuk}}, \bibinfo
  {author} {\bibfnamefont {K.}~\bibnamefont {Lenz}}, \bibinfo {author}
  {\bibfnamefont {C.}~\bibnamefont {Ludwig}}, \bibinfo {author} {\bibfnamefont
  {Y.}~\bibnamefont {Sobolev}}, \bibinfo {author} {\bibfnamefont
  {K.}~\bibnamefont {Tullney}}, \bibinfo {author} {\bibfnamefont
  {M.}~\bibnamefont {Burghoff}}, \bibinfo {author} {\bibfnamefont
  {W.}~\bibnamefont {Kilian}}, \bibinfo {author} {\bibfnamefont
  {S.}~\bibnamefont {Knappe-Gr{\"u}neberg}}, \bibinfo {author} {\bibfnamefont
  {W.}~\bibnamefont {M{\"u}ller}}, \bibinfo {author} {\bibfnamefont
  {A.}~\bibnamefont {Schnabel}}, \bibinfo {author} {\bibfnamefont
  {F.}~\bibnamefont {Seifert}}, \bibinfo {author} {\bibfnamefont
  {L.}~\bibnamefont {Trahms}}, \ and\ \bibinfo {author} {\bibfnamefont
  {S.}~\bibnamefont {Bae{\ss}ler}},\ }\href {\doibase
  10.1140/epjd/e2010-00044-5} {\bibfield  {journal} {\bibinfo  {journal} {The
  European Physical Journal D}\ }\textbf {\bibinfo {volume} {57}},\ \bibinfo
  {pages} {303} (\bibinfo {year} {2010})}\BibitemShut {NoStop}%
\bibitem [{\citenamefont {Koch}\ \emph {et~al.}(2015)\citenamefont {Koch},
  \citenamefont {Bison}, \citenamefont {Gruji{\'{c}}}, \citenamefont {Heil},
  \citenamefont {Kasprzak}, \citenamefont {Knowles}, \citenamefont {Kraft},
  \citenamefont {Pazgalev}, \citenamefont {Schnabel}, \citenamefont {Voigt},\
  and\ \citenamefont {Weis}}]{Koch2015}%
  \BibitemOpen
  \bibfield  {author} {\bibinfo {author} {\bibfnamefont {H.-C.}\ \bibnamefont
  {Koch}}, \bibinfo {author} {\bibfnamefont {G.}~\bibnamefont {Bison}},
  \bibinfo {author} {\bibfnamefont {Z.~D.}\ \bibnamefont {Gruji{\'{c}}}},
  \bibinfo {author} {\bibfnamefont {W.}~\bibnamefont {Heil}}, \bibinfo {author}
  {\bibfnamefont {M.}~\bibnamefont {Kasprzak}}, \bibinfo {author}
  {\bibfnamefont {P.}~\bibnamefont {Knowles}}, \bibinfo {author} {\bibfnamefont
  {A.}~\bibnamefont {Kraft}}, \bibinfo {author} {\bibfnamefont
  {A.}~\bibnamefont {Pazgalev}}, \bibinfo {author} {\bibfnamefont
  {A.}~\bibnamefont {Schnabel}}, \bibinfo {author} {\bibfnamefont
  {J.}~\bibnamefont {Voigt}}, \ and\ \bibinfo {author} {\bibfnamefont
  {A.}~\bibnamefont {Weis}},\ }\href {\doibase 10.1140/epjd/e2015-60018-7}
  {\bibfield  {journal} {\bibinfo  {journal} {The European Physical Journal D}\
  }\textbf {\bibinfo {volume} {69}},\ \bibinfo {pages} {202} (\bibinfo {year}
  {2015})}\BibitemShut {NoStop}%
\bibitem [{\citenamefont {Hunter}\ \emph {et~al.}(2018)\citenamefont {Hunter},
  \citenamefont {Piccolomo}, \citenamefont {Pritchard}, \citenamefont
  {Brockie}, \citenamefont {Dyer},\ and\ \citenamefont {Riis}}]{Hunter2018}%
  \BibitemOpen
  \bibfield  {author} {\bibinfo {author} {\bibfnamefont {D.}~\bibnamefont
  {Hunter}}, \bibinfo {author} {\bibfnamefont {S.}~\bibnamefont {Piccolomo}},
  \bibinfo {author} {\bibfnamefont {J.~D.}\ \bibnamefont {Pritchard}}, \bibinfo
  {author} {\bibfnamefont {N.~L.}\ \bibnamefont {Brockie}}, \bibinfo {author}
  {\bibfnamefont {T.~E.}\ \bibnamefont {Dyer}}, \ and\ \bibinfo {author}
  {\bibfnamefont {E.}~\bibnamefont {Riis}},\ }\href {\doibase
  10.1103/PhysRevApplied.10.014002} {\bibfield  {journal} {\bibinfo  {journal}
  {Phys. Rev. Applied}\ }\textbf {\bibinfo {volume} {10}},\ \bibinfo {pages}
  {014002} (\bibinfo {year} {2018})}\BibitemShut {NoStop}%
\bibitem [{\citenamefont {Huang}\ and\ \citenamefont
  {Lehmann}(2013)}]{Huang2013}%
  \BibitemOpen
  \bibfield  {author} {\bibinfo {author} {\bibfnamefont {H.}~\bibnamefont
  {Huang}}\ and\ \bibinfo {author} {\bibfnamefont {K.~K.}\ \bibnamefont
  {Lehmann}},\ }\href {\doibase 10.1021/jp406691e} {\bibfield  {journal}
  {\bibinfo  {journal} {The Journal of Physical Chemistry A}\ }\textbf
  {\bibinfo {volume} {117}},\ \bibinfo {pages} {13399} (\bibinfo {year}
  {2013})}\BibitemShut {NoStop}%
\bibitem [{\citenamefont {Westberg}\ and\ \citenamefont
  {Wysocki}(2017)}]{Westberg2017}%
  \BibitemOpen
  \bibfield  {author} {\bibinfo {author} {\bibfnamefont {J.}~\bibnamefont
  {Westberg}}\ and\ \bibinfo {author} {\bibfnamefont {G.}~\bibnamefont
  {Wysocki}},\ }\href {\doibase 10.1007/s00340-017-6743-6} {\bibfield
  {journal} {\bibinfo  {journal} {Applied Physics B}\ }\textbf {\bibinfo
  {volume} {123}},\ \bibinfo {pages} {168} (\bibinfo {year}
  {2017})}\BibitemShut {NoStop}%
\bibitem [{\citenamefont {Morville}\ \emph {et~al.}(2005)\citenamefont
  {Morville}, \citenamefont {Kassi}, \citenamefont {Chenevier},\ and\
  \citenamefont {Romanini}}]{Morville2005}%
  \BibitemOpen
  \bibfield  {author} {\bibinfo {author} {\bibfnamefont {J.}~\bibnamefont
  {Morville}}, \bibinfo {author} {\bibfnamefont {S.}~\bibnamefont {Kassi}},
  \bibinfo {author} {\bibfnamefont {M.}~\bibnamefont {Chenevier}}, \ and\
  \bibinfo {author} {\bibfnamefont {D.}~\bibnamefont {Romanini}},\ }\href
  {\doibase 10.1007/s00340-005-1828-z} {\bibfield  {journal} {\bibinfo
  {journal} {Applied Physics B}\ }\textbf {\bibinfo {volume} {80}},\ \bibinfo
  {pages} {1027} (\bibinfo {year} {2005})}\BibitemShut {NoStop}%
\bibitem [{\citenamefont {Burkart}, \citenamefont {Romanini},\ and\
  \citenamefont {Kassi}(2014)}]{Burkart2014}%
  \BibitemOpen
  \bibfield  {author} {\bibinfo {author} {\bibfnamefont {J.}~\bibnamefont
  {Burkart}}, \bibinfo {author} {\bibfnamefont {D.}~\bibnamefont {Romanini}}, \
  and\ \bibinfo {author} {\bibfnamefont {S.}~\bibnamefont {Kassi}},\ }\href
  {\doibase 10.1364/OL.39.004695} {\bibfield  {journal} {\bibinfo  {journal}
  {Opt. Lett.}\ }\textbf {\bibinfo {volume} {39}},\ \bibinfo {pages} {4695}
  (\bibinfo {year} {2014})}\BibitemShut {NoStop}%
\bibitem [{\citenamefont {Wang}\ \emph {et~al.}(2015)\citenamefont {Wang},
  \citenamefont {Nikodem}, \citenamefont {Zhang}, \citenamefont {Cikach},
  \citenamefont {Barnes}, \citenamefont {Comhair}, \citenamefont {Dweik},
  \citenamefont {Kao},\ and\ \citenamefont {Wysocki}}]{Wang2015}%
  \BibitemOpen
  \bibfield  {author} {\bibinfo {author} {\bibfnamefont {Y.}~\bibnamefont
  {Wang}}, \bibinfo {author} {\bibfnamefont {M.}~\bibnamefont {Nikodem}},
  \bibinfo {author} {\bibfnamefont {E.}~\bibnamefont {Zhang}}, \bibinfo
  {author} {\bibfnamefont {F.}~\bibnamefont {Cikach}}, \bibinfo {author}
  {\bibfnamefont {J.}~\bibnamefont {Barnes}}, \bibinfo {author} {\bibfnamefont
  {S.}~\bibnamefont {Comhair}}, \bibinfo {author} {\bibfnamefont {R.~A.}\
  \bibnamefont {Dweik}}, \bibinfo {author} {\bibfnamefont {C.}~\bibnamefont
  {Kao}}, \ and\ \bibinfo {author} {\bibfnamefont {G.}~\bibnamefont
  {Wysocki}},\ }\href {https://doi.org/10.1038/srep09096
  http://10.0.4.14/srep09096
  https://www.nature.com/articles/srep09096{\#}supplementary-information}
  {\bibfield  {journal} {\bibinfo  {journal} {Scientific Reports}\ }\textbf
  {\bibinfo {volume} {5}},\ \bibinfo {pages} {9096} (\bibinfo {year}
  {2015})}\BibitemShut {NoStop}%
\bibitem [{\citenamefont {Henderson}\ \emph {et~al.}(2018)\citenamefont
  {Henderson}, \citenamefont {Khodabakhsh}, \citenamefont {Mets{\"a}l{\"a}},
  \citenamefont {Ventrillard}, \citenamefont {Schmidt}, \citenamefont
  {Romanini}, \citenamefont {Ritchie}, \citenamefont {te~Lintel~Hekkert},
  \citenamefont {Briot}, \citenamefont {Risby}, \citenamefont {Marczin},
  \citenamefont {Harren},\ and\ \citenamefont {Cristescu}}]{Henderson2018}%
  \BibitemOpen
  \bibfield  {author} {\bibinfo {author} {\bibfnamefont {B.}~\bibnamefont
  {Henderson}}, \bibinfo {author} {\bibfnamefont {A.}~\bibnamefont
  {Khodabakhsh}}, \bibinfo {author} {\bibfnamefont {M.}~\bibnamefont
  {Mets{\"a}l{\"a}}}, \bibinfo {author} {\bibfnamefont {I.}~\bibnamefont
  {Ventrillard}}, \bibinfo {author} {\bibfnamefont {F.~M.}\ \bibnamefont
  {Schmidt}}, \bibinfo {author} {\bibfnamefont {D.}~\bibnamefont {Romanini}},
  \bibinfo {author} {\bibfnamefont {G.~A.~D.}\ \bibnamefont {Ritchie}},
  \bibinfo {author} {\bibfnamefont {S.}~\bibnamefont {te~Lintel~Hekkert}},
  \bibinfo {author} {\bibfnamefont {R.}~\bibnamefont {Briot}}, \bibinfo
  {author} {\bibfnamefont {T.}~\bibnamefont {Risby}}, \bibinfo {author}
  {\bibfnamefont {N.}~\bibnamefont {Marczin}}, \bibinfo {author} {\bibfnamefont
  {F.~J.~M.}\ \bibnamefont {Harren}}, \ and\ \bibinfo {author} {\bibfnamefont
  {S.~M.}\ \bibnamefont {Cristescu}},\ }\href {\doibase
  10.1007/s00340-018-7030-x} {\bibfield  {journal} {\bibinfo  {journal}
  {Applied Physics B}\ }\textbf {\bibinfo {volume} {124}},\ \bibinfo {pages}
  {161} (\bibinfo {year} {2018})}\BibitemShut {NoStop}%
\end{thebibliography}%
\end{document}